\documentclass{aa}

\usepackage{txfonts}
\usepackage{graphicx}
\usepackage{natbib}
\bibpunct{(}{)}{;}{a}{}{,}

\begin{document}

\title{J-PLUS: Systematic impact of metallicity on photometric calibration with the stellar locus}

\author{C.~L\'opez-Sanjuan\inst{1}
%%%%%%%%% People that contributed directly to the paper
\and H.~Yuan\inst{2}
\and H.~V\'azquez Rami\'o\inst{1}
\and J.~Varela\inst{1}
\and D.~Crist\'obal-Hornillos\inst{3}
\and P.~-E.~Tremblay\inst{4}
\and A.~Mar\'{\i}n-Franch\inst{1}
\and A.~J.~Cenarro\inst{1}
\and A.~Ederoclite\inst{5}
%%%%%%%% People that send comments
\and E.~J.~Alfaro\inst{6}
\and A.~Alvarez-Candal\inst{6,7,8}
\and S.~Daflon\inst{8}
\and A.~Hern\'an-Caballero\inst{3}
\and C.~Hern\'andez-Monteagudo\inst{9,10,1}
\and F.~M.~Jim\'enez-Esteban\inst{11,12}
\and V.~M.~Placco\inst{13}
\and E.~Tempel\inst{14}
%%%%%%%% Other builders
\and J.~Alcaniz\inst{8}
\and R.~E.~Angulo\inst{15,16}
\and R.~A.~Dupke\inst{8,17,18}
\and M.~Moles\inst{3}
\and L.~Sodr\'e Jr.\inst{5}
}

\institute{Centro de Estudios de F\'{\i}sica del Cosmos de Arag\'on (CEFCA), Unidad Asociada al CSIC, Plaza San Juan 1, 44001 Teruel, Spain\\\email{clsj@cefca.es}
        %2
        \and
        Department of Astronomy, Beijing Normal University, Beijing 100875, People's Republic of China
        %3
        \and
        Centro de Estudios de F\'{\i}sica del Cosmos de Arag\'on (CEFCA), Plaza San Juan 1, 44001 Teruel, Spain
        %4
        \and
        Department of Physics, University of Warwick, Coventry, CV4 7AL, UK
        %5
        \and
        Instituto de Astronomia, Geof\'{\i}sica e Ci\^encias Atmosf\'ericas, Universidade de S\~ao Paulo, 05508-090 S\~ao Paulo, Brazil
        %6
        \and
        Instituto de Astrof\'{\i}sica de Andaluc\'{\i}a (IAA-CSIC), Glorieta de la astronomía s/n, 18008 Granada, Spain
        %7
        \and
        IUFACyT, Universidad de Alicante, San Vicent del Raspeig, 03080 Alicante, Spain
        %8
        \and
        Observat\'orio Nacional, Rua General Jos\'e Cristino, 77 - Bairro Imperial de S\~ao Crist\'ov\~ao, 20921-400 Rio de Janeiro, Brazil
        %9
        \and
        Instituto de Astrof\'{\i}sica de Canarias (IAC), 38205 La Laguna, Spain
        %10
        \and
        Departamento de Astrof\'{\i}sica, Universidad de La Laguna (ULL), 38200 La Laguna, Spain
        %11
        \and
        Centro de Astrobiolog\'{\i}a (CSIC-INTA), ESAC Campus, Camino Bajo del Castillo s/n, 28692 Villanueva de la Ca\~nada, Spain
        %12
        \and
        Spanish Virtual Observatory, 28692 Villanueva de la Ca\~nada, Spain
        %13
        \and
        Community Science and Data Center/NSF’s NOIRLab, 950 N. Cherry Ave., Tucson, AZ 85719, USA
        %14
        \and
        Tartu Observatory, University of Tartu, Observatooriumi 1, 61602 T\~oravere, Estonia
        %15
        \and
        Donostia International Physics Centre (DIPC), Paseo Manuel de Lardizabal 4, 20018 Donostia-San Sebastián, Spain
        %16
        \and
        IKERBASQUE, Basque Foundation for Science, 48013, Bilbao, Spain
        %17
        \and
        University of Michigan, Department of Astronomy, 1085 South University Ave., Ann Arbor, MI 48109, USA
        %18
        \and
        University of Alabama, Department of Physics and Astronomy, Gallalee Hall, Tuscaloosa, AL 35401, USA
}

\date{Received 29 January 2021 / Accepted 13 July 2021}

\abstract
{}
{We present the photometric calibration of the twelve optical passbands for the Javalambre Photometric Local Universe Survey (J-PLUS) second data release (DR2), comprising $1\,088$ pointings of two square degrees, and study the systematic impact of metallicity on the stellar locus technique.}
{The [Fe/H] metallicity from the Large Sky Area Multi-Object Fiber Spectroscopic Telescope (LAMOST) for $146\,184$ high-quality calibration stars, defined with signal-to-noise ratio larger than ten in J-PLUS passbands and larger than three in {\it Gaia} parallax, was used to compute the metallicity-dependent stellar locus (ZSL). The initial homogenization of J-PLUS photometry, performed with a unique stellar locus, was refined by including the metallicity effect in colors via the ZSL.}
{The variation of the average metallicity along the Milky Way produces a systematic offset in J-PLUS calibration. This effect is well above 1\% for the bluer passbands and amounts $0.07$, $0.07$, $0.05$, $0.03$, and $0.02$ mag in $u$, $J0378$, $J0395$, $J0410$, and $J0430$, respectively. We modeled this effect with the Milky Way location of the J-PLUS pointing, also providing an updated calibration for those observations without LAMOST information. The estimated accuracy in the calibration after including the metallicity effect is at 1\% for the bluer J-PLUS passbands and below for the rest.}
{Photometric calibration with the stellar locus technique is prone to significant systematic bias in the Milky Way for passbands bluer than $\lambda = 4\,500$ \AA. The calibration method for J-PLUS DR2 reaches 1-2\% precision and 1\% accuracy for 12 optical filters within an area of $2\,176$ square degrees.}


\titlerunning{J-PLUS. Systematic impact of metallicity on photometric calibration with the stellar locus}

\authorrunning{L\'opez-Sanjuan et al.}

\maketitle

\section{Introduction}\label{sec:intro}
One fundamental step in the data processing of any imaging survey is the photometric calibration of the observations. The calibration process aims to translate the observed counts in astronomical images to a physical flux scale referred to the top of the atmosphere. Because accurate colors are needed to derive photometric redshifts for galaxies, atmospheric parameters for Milky Way (MW) stars, and surface characteristics for minor bodies; and reliable absolute fluxes are involved in the estimation of the luminosity and the stellar mass of galaxies, current and future photometric surveys target a calibration uncertainty at the 1\% level and below to reach their ambitious scientific goals.

\begin{table*} 
\caption{J-PLUS photometric system.}
\label{tab:JPLUS_filters}
\centering 
        \begin{tabular}{l c c c c l }
        \hline\hline\rule{0pt}{3ex} 
        Passband $(\mathcal{X})$   & Central wavelength    & FWHM  & $m_{\rm lim}^{\rm DR2}$  & $k_{\mathcal{X}} = \frac{A_{\mathcal{X}}}{E(B-V)}$ & Comments\\\rule{0pt}{2ex} 
                &   [nm]                & [nm]           &  [AB]\tablefootmark{a}   &        &  \\
        \hline\rule{0pt}{2ex}
        $u$             &348.5  &50.8           &       20.8    &  4.479  & In common with J-PAS\\ 
        $J0378$         &378.5  &16.8           &       20.8    &  4.294  & [OII]; in common with J-PAS\\ 
        $J0395$         &395.0  &10.0           &       20.8    &  4.226  & Ca H$+$K; similar to the CaHK filter from {\it Pristine}\\ 
        $J0410$         &410.0  &20.0           &       21.0    &  4.023  & H$_\delta$\\ 
        $J0430$         &430.0  &20.0           &       21.0    &  3.859  & G band\\ 
        $g$             &480.3  &140.9          &       21.8    &  3.398  & SDSS\\ 
        $J0515$         &515.0  &20.0           &       21.0    &  3.148  & Mg$b$ Triplet\\ 
        $r$             &625.4  &138.8          &       21.8    &  2.383  & SDSS\\ 
        $J0660$         &660.0  &13.8           &       21.0    &  2.161  & H$\alpha$; in common with J-PAS\\ 
        $i$             &766.8  &153.5          &       21.3    &  1.743  & SDSS\\ 
        $J0861$         &861.0  &40.0           &       20.4    &  1.381  & Ca Triplet\\ 
        $z$             &911.4  &140.9          &       20.5    &  1.289  & SDSS\\ 
        \hline 
\end{tabular}
\tablefoot{
\tablefoottext{a} {Limiting magnitude (5$\sigma$, 3 arcsec diameter aperture) of J-PLUS DR2.}
}

\end{table*}

One particular approach to performing the photometric calibration is the
use of the stellar locus \citep{covey07, high09, kelly14}. This procedure takes advantage of the way stars with different stellar parameters populate color-color diagrams, defining a well-constrained region (stellar locus) whose shape depends on the specific colors used. The match between the instrumental data and a reference stellar locus provides the flux calibration of the images.

The stellar locus technique is able to provide a photometric calibration without the need for dedicated calibration images of standard stars, saving telescope time and optimizing operations. Its main assumption is that the reference locus is valid for any observed position. In the general case, the stellar photometry in the Galaxy is affected by the amount of interstellar matter that star-light passes through before reaching the observer and by possible local variations in the extinction law. This leaves two solutions to define the reference locus: de-reddening the photometry or choosing a set of dust-free objects \citep{high09}.

In addition to the interstellar extinction, the average properties of the stars also vary with their position in the MW. The stellar locus location for main-sequence (MS) stars is dominated by temperature variations, so the measured correlation in color-color diagrams is roughly a temperature sequence. However, the metallicity is also a relevant parameter that affects the stellar locus location appreciably \citep[e.g.,][]{lenz98,ivezic08,yuan15,kesseli17}, especially at the bluer optical passbands. With the average metallicity of the observed MW stars decreasing as we move from disk-dominated ([Fe/H] $\sim -0.5$ dex) to halo-dominated  ([Fe/H] $\sim -1.5$ dex) sky positions \citep{ivezic08}, the location of the stellar locus changes accordingly, and the assumption of a position-independent reference locus is not valid. This metallicity effect is a source of systematic in the calibration with the stellar locus technique.  

Several large-area photometric surveys covering the blue edge ($\lambda < 4\,500$ \AA) of the optical range rely on the stellar locus technique for calibration. We highlight the Kilo-Degree Survey (KiDS, \citealt{kids_dr4}; $ugriz$ broad bands), the {\it Pristine} survey (\citealt{pristine}; a unique CaHK filter of 98 \AA\ width centered at 3952 \AA), the SkyMapper Southern Survey (SMSS, \citealt{skymapper_dr1, skymapper_dr2}; $uvgriz$ passbands), and the Javalambre Photometric Local Universe Survey (J-PLUS, \citealt{cenarro19}; 5 broad + 7 medium optical filters as summarized in Table~\ref{tab:JPLUS_filters}). There are hints in these surveys about the impact of metallicity variations in the stellar locus calibration. For example, the comparison of the KiDS $u$ band with Sloan Digital Sky Survey (SDSS, \citealt{sdssdr8}) photometry reveals a systematic variation with Galactic latitude, which the authors link to the change in metallicity \citep{kids_dr4}. Furthermore, the photometric metallicity derived from {\it Pristine} presents a systematic variation with the sky position when the stellar locus calibration is performed with $0.4 < (g - i) < 1.2$ stars. They conclude that this is a reflection of the metallicity impact on the stellar locus, and proper measurements are achieved by calibrating with dwarf MS stars in the $1.2 < (g - i) < 2.4$ color range \citep{pristine}.

The J-PLUS second data release (DR2; \citealt{jplus_dr2}), covering $2\,176$ deg$^2$, was made public in November 2020, and we describe its photometric calibration here. It is based on the stellar and white dwarf loci procedure detailed in \citet{clsj19jcal}, which was applied to the J-PLUS first data release (DR1). For the present work, we took advantage of the [Fe/H] information provided by the Large Sky Area Multi-Object Fiber Spectroscopic Telescope (LAMOST, \citealt{lamost}) surveys to implement the metallicity-dependent stellar locus for calibration. That improved the accuracy of the published J-PLUS DR2 photometry with respect to previous releases, especially at passbands bluer than $\lambda = 4\,500$ \AA, and it highlights the systematic variation of the photometric solution with the position in the sky when metallicity effects are neglected.

In addition to a metallicity-dependent stellar locus, the access to massive spectroscopic information also permits the application of the stellar color regression (SCR, \citealt{scr,huang21}) method. Using $T_{\rm eff}$, $\log g$, and [Fe/H] from spectroscopy, the SCR matches stars of the same properties (i.e. intrinsic colors) and assigns observed color differences to the effect of interstellar extinction. This permits the homogenization of the photometric solution by naturally accounting for temperature, gravity, metallicity, and extinction effects. The application of the SCR to J-PLUS data is beyond the scope of the present paper, and it is explored in a forthcoming work.

This paper is organized as follows. The J-PLUS DR2 and the ancillary data used on its calibration are presented in Sect.~\ref{sec:data}. The calibration methodology is summarized in Sect.~\ref{sec:method}, highlighting the addition of the metallicity-dependent stellar locus in the process. The precision, accuracy, and the systematic impact of metallicity in the J-PLUS DR2 calibration are discussed in Sect.~\ref{sect:discussion}. We devote Sect.~\ref{sec:summary} to summarizing this work and presenting our conclusions. Magnitudes are given in the AB system \citep{oke83}.

\section{Data}\label{sec:data}
\subsection{J-PLUS photometric data}\label{sec:jplus}
J-PLUS\footnote{\url{www.j-plus.es}} is being conducted from the Observatorio Astrof\'{\i}sico de Javalambre (OAJ, Teruel, Spain; \citealt{oaj}) using the 83\,cm Javalambre Auxiliary Survey Telescope (JAST80) and T80Cam, a panoramic camera of 9.2k $\times$ 9.2k pixels that provides a $2\deg^2$ field of view (FoV) with a pixel scale of 0.55$^{\prime\prime}$pix$^{-1}$ \citep{t80cam}. The J-PLUS filter system, composed of twelve bands, is summarized in Table~\ref{tab:JPLUS_filters}. These filters were designed to optimize the characterization of MW stars. The J-PLUS observational strategy, image reduction, and main scientific goals are presented in \citet{cenarro19}. In addition to its scientific potential, J-PLUS was defined with the technical goal of ensuring the photometric calibration of the Javalambre Physics of the Accelerating Universe Astrophysical Survey (J-PAS; \citealt{jpas, minijpas}), which will scan thousands of square degrees with 56 narrow bands of $\sim 140\ $ \AA{} width down to $m \sim 22.5$ mag from the OAJ. 

The J-PLUS DR2 comprises $1\,088$ pointings ($2\,176$ deg$^2$) observed and reduced in all survey bands \citep{jplus_dr2}. The limiting magnitudes (5$\sigma$, 3$^{\prime\prime}$ aperture) of the DR2 are presented in Table~\ref{tab:JPLUS_filters} for reference. The median point spread function (PSF) full width at half maximum (FWHM) in the DR2 $r$-band images is 1.1$^{\prime\prime}$. Source detection was done in the $r$ band using \texttt{SExtractor} \citep{sextractor}, and the flux measurement in the twelve J-PLUS bands was performed at the position of the detected sources using the aperture defined in the $r$-band image. Objects near the borders of the images, close to bright stars or affected by optical artifacts, were masked. The DR2 is publicly available at the J-PLUS website\footnote{\url{www.j-plus.es/datareleases/data_release_dr2}}. 

We highlight that the published J-PLUS DR2 photometry already includes all the calibration steps presented in Sect.~\ref{sec:method}. In addition to J-PLUS information, ancillary data from {\it Gaia}, the Panoramic Survey Telescope and Rapid Response System (Pan-STARRS), and LAMOST was used in the calibration process. We describe these datasets in the following.

\subsection{Pan-STARRS DR1}\label{sec:ps1}
The Pan-STARRS1 is a 1.8 m optical and near-infrared telescope located on Mount Haleakala, Hawaii. The telescope is equipped with the Gigapixel Camera 1 (GPC1), consisting of an array of 60 CCD detectors, each $4\,800 \times 4\,800$ pixels \citep{chambers16}. 

The 3$\pi$ Steradian Survey (hereafter PS1; \citealt{chambers16}) covers the sky north of declination $\delta  = -30^{\circ}$ in four SDSS-like passbands, $griz$, with an additional passband in the near-infrared, $y$. The entire filter set spans the $400 - 1\,000$ nm range \citep{tonry12}.

Astrometry and photometry were extracted by the Pan-STARRS1 Image Processing Pipeline \citep{ps1pipe,ps1cal,ps1phot,ps1pix}. PS1 photometry features a uniform flux calibration, achieving better than 1\% precision over the sky \citep{ps1cal,chambers16}. In single-epoch photometry, PS1 reaches typical 5$\sigma$ depths of 22.0, 21.8, 21.5, 20.9, and 19.7 in $grizy$, respectively \citep{chambers16}. The PS1 DR1 occurred in December 2016, and provided a static-sky catalog, stacked images from the 3$\pi$ Steradian Survey, and other data products \citep{ps1data}.

Because of its large footprint, homogeneous depth, and excellent internal calibration, PS1 photometry provides an ideal reference for the calibration of the $gri$ J-PLUS broad bands. The $z-$band photometry from PS1 was reserved to test the calibration procedure.

\subsection{Gaia DR2}\label{sec:gaia}
The {\it Gaia} spacecraft is mapping the 3D positions and kinematics of a representative fraction of MW stars \citep{gaia}. The mission will eventually provide astrometry (positions, proper motions, and parallaxes) and optical spectro-photometry for over a billion stars, as well as radial velocity measurements of more than $100$ million stars.

In the present work, we used the {\it Gaia} DR2 \citep{gaiadr2}, which is based on 22 months of observations. It contains five-parameter astrometric determinations and provides integrated photometry in three broad bands, $G$ ($330 - 1\,050$ nm), $G_{\rm BP}$ ($330 - 680$ nm), and $G_{\rm RP}$ ($630 - 1\,050$ nm), for 1.4 billion sources with $G < 21$. The typical uncertainties in {\it Gaia} DR2 measurements at $G = 17$ are $\sim0.1$ mas in parallax, $\sim2$ mmag in $G$-band photometry, and $\sim10$ mmag in $G_{\rm BP}$ and $G_{\rm RP}$ magnitudes \citep{gaiadr2}.

\subsection{LAMOST DR5}\label{sec:lamost}
LAMOST is a four-meter quasi-meridian reflecting Schmidt telescope equipped with thousands of fibers distributed in a FoV of about 20 deg$^2$. It can simultaneously collect spectra per exposure of up to $4\,000$ objects, covering the $380 - 900$ nm wavelength range at a resolving power of $R \sim 1\,800$ \citep{lamost}. The five-year Phase I LAMOST regular surveys started in the fall of 2012 and were completed in the summer of 2017. The scientific motivations and target selections of these surveys are described in \citet{zhao12}, \citet{legue}, \citet{lamost_dr1}, and \citet{lss-gac}.

In the final data release of the LAMOST Phase I surveys, the LAMOST DR5 provides $9\,027\,634$ optical spectra to the community, of which more than $90$ percent are stellar. The LAMOST DR5 provides stellar classifications and radial velocity measurements for these spectra. In the present work, we were restricted to the $5\,348\,712$ objects in the ``A, F, G, and K type star catalog''\footnote{\url{http://dr5.lamost.org/v3/doc/data-production-description#toc_16}.} This includes the basic stellar parameters $T_{\rm eff}$, $\log {\rm g}$, and [Fe/H] derived with the LAMOST stellar parameter pipeline (LASP; \citealt{wu11,lamost_dr1}).

\section{Photometric calibration of J-PLUS DR2}\label{sec:method}
The photometric calibration of the J-PLUS DR2 data follows the main steps presented in \citet{clsj19jcal} for the calibration of J-PLUS DR1. We provide a brief summary of the process in Sect.~\ref{sec:calibration}. The main improvement with respect to DR1 is the inclusion of the metallicity effect in the stellar locus location, as detailed in Sect.~\ref{sec:ZSL}.

\subsection{Definition of the zero point}
The goal of any calibration strategy is to obtain the zero point (ZP) of the observation; that relates the magnitude of the sources in passband $\mathcal{X}$ on top of the atmosphere with the magnitudes obtained from the analogue-to-digital unit (ADU) counts of the reduced images. We simplify the notation in the following using the passband name as the magnitude in such filter. Thus,
\begin{equation}
\mathcal{X} = -2.5\log_{10}({\rm ADU}_{\mathcal{X}}) + {\rm ZP}_{\mathcal{X}}.
\end{equation}
In the estimation of the J-PLUS DR2 raw catalogs, the reduced images were normalized to a one-second exposure and an arbitrary instrumental zero point ${\rm ZP}_{\mathcal{X}} = 25$ was set. This defined the instrumental magnitudes $\mathcal{X}_{\rm ins}$.

\begin{figure}[t]
\centering
\resizebox{\hsize}{!}{\includegraphics{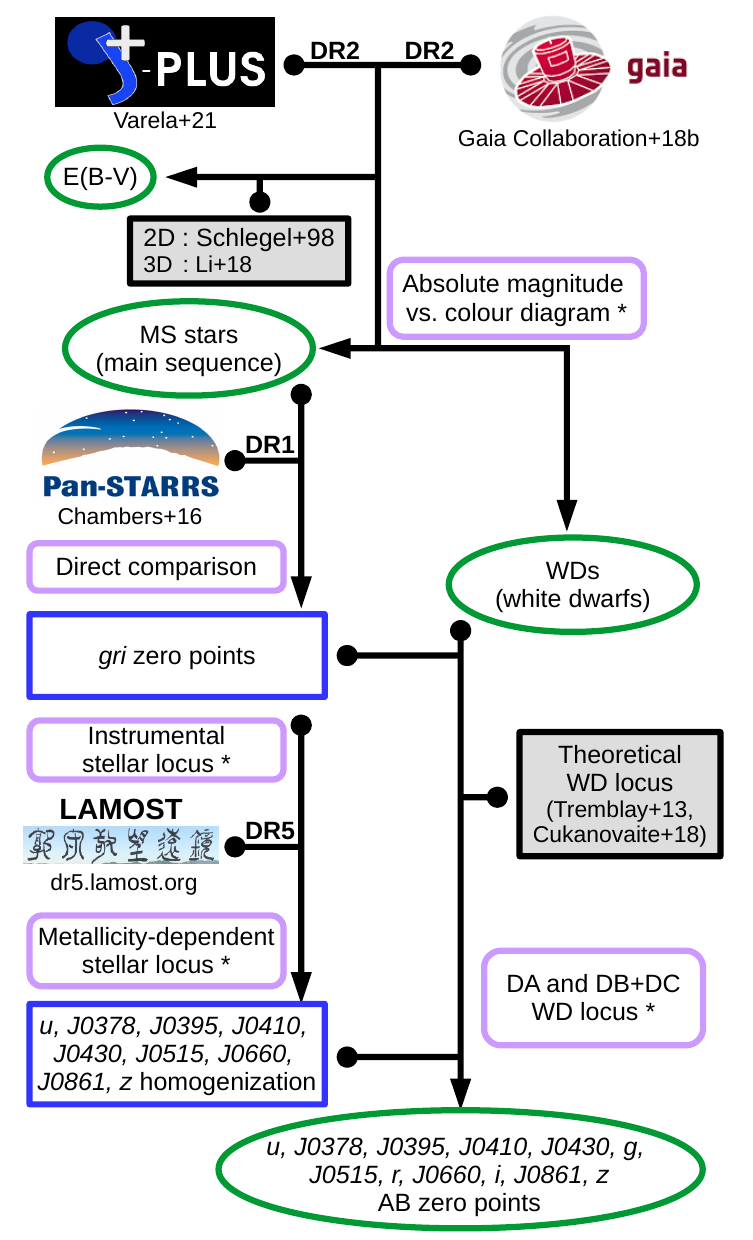}}
\caption{Updated flowchart of the calibration method used in J-PLUS DR2. Arrows that originate in small dots indicate that the preceding data product is an input to the subsequent analysis. Datasets are shown with their project logo, and external codes or models are denoted with gray boxes. The rounded boxes show the calibration steps. The asterisks indicate those steps based on dust de-reddened magnitudes. The white boxes show intermediate data products, and ovals highlight data products of the calibration process. The changes with respect to J-PLUS DR1 calibration are the modification in the assumed dust extinction and the addition of the metallicity-dependent stellar locus step in the homogenization (Sect.~\ref{sec:ZSL}).} 
\label{fig:calib_chart}
\end{figure}

The calibration process applied in J-PLUS DR2 has different steps, as described in Sect.~\ref{sec:calibration}. At the end, we estimated the zero point of the passband $\mathcal{X}$ in the pointing $p_{\rm id}$ as
\begin{align}\label{eq:zp} 
{\rm ZP} & _{\mathcal{X}}\,(p_{\rm id},X,Y) = \nonumber \\ & \Delta \mathcal{X}_{\rm atm}\,(p_{\rm id}) + P_{\mathcal{X}}\,(p_{\rm id},X,Y) + \Delta \mathcal{X}_{\rm FeH}\,(p_{\rm id}) + \Delta \mathcal{X}_{\rm WD} + 25,
\end{align}
where $\Delta \mathcal{X}_{\rm atm}$ is the term that accounts for the atmospheric extinction at the moment of the observation, $P_{\mathcal{X}}$ defines a plane that accounts for the 2D variation of the calibration with the $(X,Y)$ position of the sources on the CCD, $\Delta \mathcal{X}_{\rm FeH}$ includes the effect of the metallicity in the stellar locus homogenization process, and $\Delta \mathcal{X}_{\rm WD}$ is the global offset provided by the white dwarf (WD) locus that translates homogenized magnitudes to calibrated magnitudes outside the atmosphere. 

\subsection{Total magnitudes for calibration}
The J-PLUS instrumental magnitudes used for calibration were measured on a 6 arcsec diameter aperture. This aperture is not dominated by background noise and limits the flux contamination from neighboring sources, although it is not large enough to capture the total flux of the stars. Thus, we applied an aperture correction $C_{\rm aper}$ that depends on the pointing and the passband. 

The aperture correction was computed from the growth curve of bright, non-saturated stars in the pointing. For each star, increasingly larger circular apertures were measured until convergence within errors. This defined the aperture size that provides the total magnitude of the sources in the pointing; that is then compared with the magnitude at 6 arcsec aperture to provide $C_{\rm aper}$. The typical number of stars used is 50, and the median aperture correction varies from $C_{\rm aper} = -0.09$ mag in the $u$ band to $C_{\rm aper} = -0.11$ mag in the $z$ band, with a median value of $C_{\rm aper} = -0.09$ mag for all the filters. The typical uncertainty in the correction aperture, estimated from the dispersion in the measurements, is $\sim 2$ mmag. We assumed that the J-PLUS 6 arcsec magnitudes corrected for aperture effects provided the total flux of stars.

\subsection{Extinction correction}\label{sec:ext}
We worked with dust de-reddened magnitudes and colors in several stages of the calibration process. We empirically computed the extinction coefficients $k_{\mathcal{X}}$ of each J-PLUS passband, presented in Table~\ref{tab:JPLUS_filters}, by applying the star-pair technique described in \citet{yuan13} to J-PLUS DR1. 

The de-reddened J-PLUS photometry, either instrumental or calibrated, is noted with the subscript $0$ and was obtained as
\begin{equation}
\mathcal{X}_0 = \mathcal{X} - k_{\mathcal{X}} E(B-V).
\end{equation}
We estimated the color excess at infinite distance of each J-PLUS source from the \citet{sfd98} extinction map, noted $E(B-V)_{\rm SFD}$. The stars used in the calibration process have distance information from {\it Gaia} DR2 parallaxes (Sect.~\ref{sec:calibration}), and we included the 3D information using the MW dust model presented in \cite{li18}. We integrated the dust model to infinity and to the distance provided by {\it Gaia} at star location, scaling the color excess from the \citet{sfd98} map accordingly to obtain $E(B-V)$. 

We tested our assumed $E(B-V)$ using the star-pair method \citep{yuan13}. We started by gathering J-PLUS stars with $E(B-V)_{\rm SFD} \leq 0.02$ and LAMOST spectroscopic data as a reference sample. We de-reddened their observed $(g-r)$ color with the assumed $E(B-V)$ and the empirical extinction coefficients in Table~\ref{tab:JPLUS_filters}. This provided a set of intrinsic colors, $(g-r)_0$. Then, for each J-PLUS star with LAMOST information and irrespectively of its color excess, we searched for those sources in the reference sample with a difference in effective temperature of $\delta T_{\rm eff} < 50$ K, in surface gravity $\delta \log {\rm g} < 0.25$ dex, and in metallicity $\delta {\rm [Fe/H]} < 0.1$ dex. Assuming a linear variation of $(g-r)_0$ in these properties, the expected intrinsic color of the target star was evaluated and its empirical color excess $E(g-r) = (g-r) - (g-r)_0$ estimated. The comparison between the assumed $E(B-V)$ and the estimated $E(g-r)$ was linear, $E(g-r) = 1.02E(B-V)$, and presented a dispersion of $0.012$ mag. We conclude that our assumed $E(B-V)$ is a proper proxy for the real color excess of the calibration stars and permits to set the uncertainty in $E(B-V)$ to $0.012$ mag. We further test the assumed extinction correction in Sect.~\ref{sect:ebvtest}.

\subsection{Scheme of the calibration process}\label{sec:calibration}
 In this section, we provide a brief summary of the steps involved in the photometric calibration of J-PLUS DR2 images. The updated flowchart of the calibration process is presented in Fig.~\ref{fig:calib_chart}. We refer the reader to \citet{clsj19jcal} for an extensive description of the calibration procedure, but the metallicity-dependent stellar locus step, added for J-PLUS DR2, is described in Sect.~\ref{sec:ZSL}. 
 
We started by defining a high-quality sample of MS stars for calibration. We selected those sources in common between J-PLUS DR2 and {\it Gaia} DR2 with signal-to-noise ratio (S/N) larger than ten in all the photometric bands and with S/N > 3 in {\it Gaia} parallax. We constructed the dust de-reddened absolute $G$ magnitude versus $G_{\rm BP} - G_{\rm RP}$ diagram and selected those sources belonging to the main sequence. This provided $1\,117\,073$ MS calibration stars, with a median of 822 calibration stars per pointing and a minimum of 92 stars.
    
Then, we calibrated the $gri$ broad-band filters with PS1 photometry. The J-PLUS instrumental magnitudes were compared with the PSF magnitudes in PS1 after accounting for the color terms between both photometric systems. This step provides $\Delta {\mathcal{X}}_{\rm atm}$ and the 2D variation along the CCD of the $gri$ broad-band filters. Because we used PS1 calibrated magnitudes as reference, $\Delta {\mathcal{X}}_{\rm FeH} = 0$ and $\Delta \mathcal{X}_{\rm WD} \sim 0$. The latter term is not zero because residual differences between J-PLUS and PS1 photometric systems can exist, as discussed in Sect.~\ref{sect:wds}.

The next step was the homogenization of the remaining passbands. The initial homogenization was performed with the instrumental stellar locus (ISL). We computed the dust de-reddened $(\mathcal{X}_{\rm ins} - r)_0$ versus $(g - i)_0$ color-color diagrams of the MS calibration stars. From these, we computed the offsets that lead to a consistent ISL among all the J-PLUS DR2 pointings. This provides $\Delta {\mathcal{X}}_{\rm atm}$ and the 2D variation along the CCD for the other nine J-PLUS passbands. After this step, we defined the ISL magnitudes as
    \begin{equation}
        \mathcal{X}_{\rm ISL} = \mathcal{X}_{\rm ins} + \Delta \mathcal{X}_{\rm atm} + P_{\mathcal{X}}.
    \end{equation}

The homogenization was refined with the metallicity-dependent stellar locus (ZSL). We updated the methodology presented in \citet{clsj19jcal} by including the effect of metallicity in the stellar locus location. We used the metallicity measurements from LAMOST DR5, and the procedure is fully detailed in Sect.~\ref{sec:ZSL}. This step provided $\Delta {\mathcal{X}}_{\rm FeH}$, defining the ISL + ZSL magnitudes:
    \begin{equation}
        \mathcal{X}_{\rm ISL+ ZSL} = \mathcal{X}_{\rm ISL} + \Delta {\mathcal{X}}_{\rm FeH}.
    \end{equation}
    
Finally, the performed the absolute color calibration with the white dwarf locus. From the {\it Gaia} absolute magnitude versus color diagram in the first step, we also selected 639 high-quality WDs. We compared the observed color-color locus in $(\mathcal{X}_{\rm ISL+ZSL} - r)_0$ versus $(g - i)_0$ with the theoretical expectations from pure hydrogen (DA; \citealt{tremblay13}) and pure helium (DB and DC; \citealt{cukanovaite18}) models. The Bayesian modeling of the WD locus provided the $\Delta \mathcal{X}_{\rm WD}$ for all the passbands except $r$, which was used as the reference band in the process. The technical details of this step are presented in Appendix~\ref{app:wdlocus}. 

This calibration process, including the metallicity-dependent stellar locus, was applied to the J-PLUS DR2 published data. We stress that the calibrated photometry is on top of the atmosphere, and it is therefore not corrected by interstellar reddening. The performance of J-PLUS DR2 calibration is presented in Sect.~\ref{sect:discussion}. The median zero points obtained after the complete calibration process are presented in Table~\ref{tab:jplus_calib} for reference.

\begin{figure}[t]
\centering
\resizebox{\hsize}{!}{\includegraphics{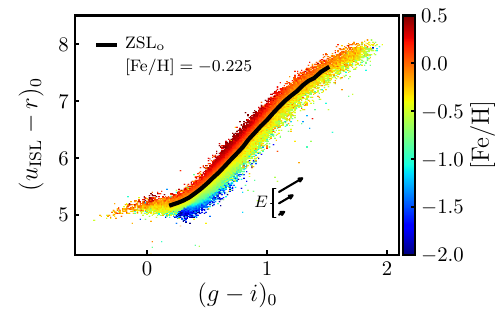}}
\resizebox{\hsize}{!}{\includegraphics{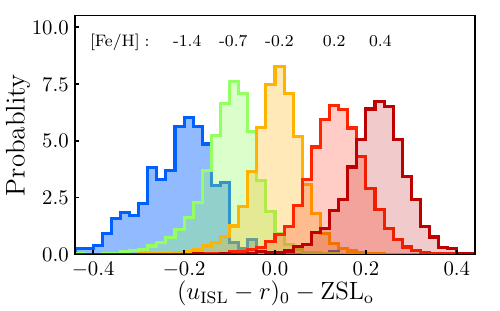}}
\caption{{\it Top panel}: Binned $(u_{\rm ISL} - r)_0$ versus $(g-i)_0$ color-color diagram. The color scale shows the median [Fe/H] in each bin estimated from LAMOST spectra. The black solid line marks the stellar locus for $-0.25 < {\rm [Fe/H]} < -0.20$ stars in the range $0.2 < (g-i)_0 < 1.5$, noted ZSL$_{\rm o}$. The arrows show the color excess vector ($E$) for the 50th, 90th, and 99th percentiles of the sources' $E(B-V)$ distributions, corresponding to $E(B-V) = 0.03, 0.08$, and $0.12$, respectively. {\it Bottom panel}: Normalized histogram of the $(u_{\rm ISL} - r)_0$ color difference with respect to ZSL$_{\rm o}$ for samples of different metallicities, defined with a central [Fe/H] $\pm\,0.1$ dex. The central metallicity of each sample is labeled in the panel.}
\label{fig:zsl_1d}
\end{figure}

\begin{figure}[t]
\centering
\resizebox{\hsize}{!}{\includegraphics{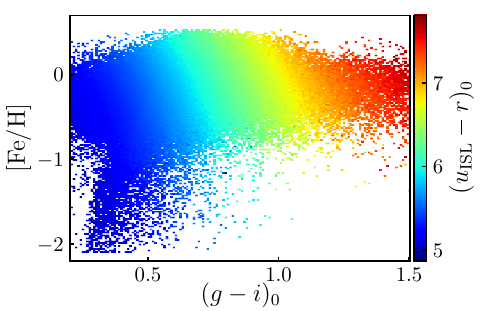}}
\resizebox{\hsize}{!}{\includegraphics{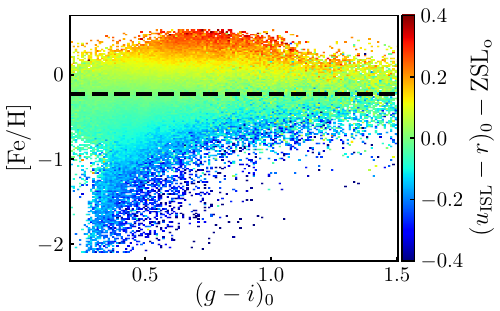}}
\caption{Binned metallicity versus $(g-i)_0$ color diagram of the MS calibration stars with measurements from LAMOST. {\it Top panel}: Mean $(u_{\rm ISL} - r)_0$ color in each bin, defining the metallicity-dependent stellar locus (ZSL). {\it Bottom panel}: $(u_{\rm ISL} - r)_0$ color difference with respect to ZSL$_{\rm o}$. The median metallicity of the reference locus is marked with the black dashed line.}
\label{fig:zsl_2d}
\end{figure}

\begin{figure}[t]
\centering
\resizebox{\hsize}{!}{\includegraphics{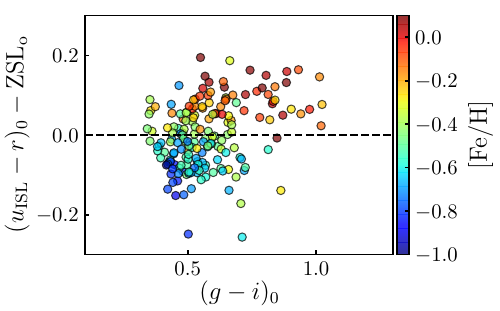}}

\resizebox{\hsize}{!}{\includegraphics{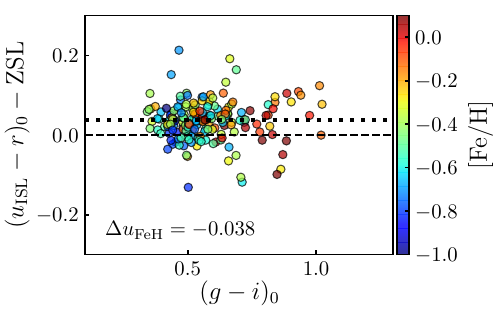}}
\caption{$(u_{\rm ISL} - r)_0$ color difference with respect to the reference locus ZSL$_{\rm o}$ ({\it top panel}) and the ZSL ({\it bottom panel}) as a function of $(g-i)_0$ for the MS calibration stars with LAMOST information in pointing $p_{\rm id} = 00066$. The colored scale in both panels shows the spectroscopic [Fe/H] from LAMOST. The dashed lines mark zero offset. The dotted line marks the median difference with respect to the ZSL. The derived metallicity offset is labeled in the {\it bottom panel}.
}
\label{fig:deltafeh}
\end{figure}

\subsection{Implementation of the metallicity-dependent stellar locus}\label{sec:ZSL}
The calibration process presented in \citet{clsj19jcal} and summarized in the previous section was updated for the published J-PLUS DR2 to include the impact of metallicity in the stellar locus location. We use the $u$ band as reference to illustrate the process, and the methodology was similar for other J-PLUS passbands except $gri$, which were anchored to PS1 photometry. The improvement in the accuracy of J-PLUS calibration along the surveyed area from this step is presented in Sect.~\ref{sec:accuracy}.

\subsubsection{LAMOST cross-match with the calibration sample}
We started by gathering the [Fe/H] (dubbed metallicity hereafter) information of the MS calibration stars. We cross-matched the calibration sample with the LAMOST catalog using a 1 arcsec radius. A total of $146\,184$ sources in common were retrieved. The median uncertainty in [Fe/H] is 0.1 dex, providing a high-quality dataset to derive the metallicity-dependent stellar locus.

Despite the large sky coverage of LAMOST, not all the J-PLUS pointings have metallicity information. We have 178 (16\%) pointings with fewer than ten calibration stars in common with LAMOST. This implies that the metallicity-dependent stellar locus procedure detailed in Sect.~\ref{sec:ZSLoff} cannot be applied to all J-PLUS DR2 observations. We circumvented this limitation by modeling the offset in the stellar locus due to metallicity with the MW location (Sect.~\ref{sec:ZSLmodel}).

\begin{figure*}[t]
\centering
\resizebox{0.49\hsize}{!}{\includegraphics{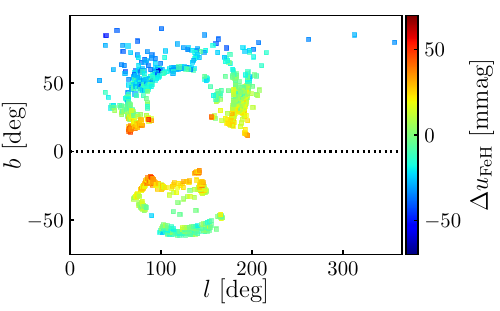}}
\resizebox{0.49\hsize}{!}{\includegraphics{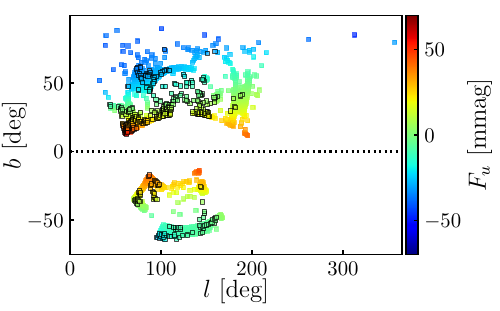}}

\resizebox{0.49\hsize}{!}{\includegraphics{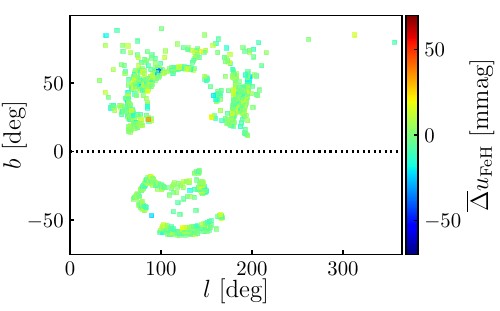}}
\resizebox{0.49\hsize}{!}{\includegraphics{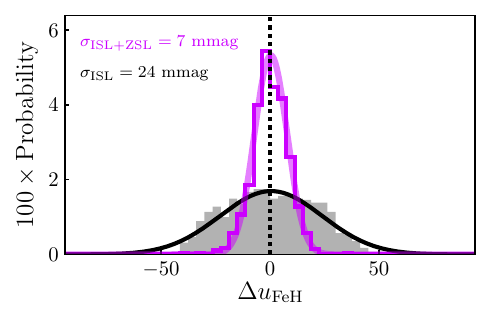}}
\caption{Photometric offset for each J-PLUS DR2 pointing estimated from the metallicity-dependent stellar locus. {\it Top left panel}: Initial offset in Galactic coordinates with the homogenization from ISL, $\Delta u_{\rm FeH}$. {\it Top right panel}: Modeled metallicity offset $F_{u}$ in Galactic coordinates. Those pointings without offset estimation and not used in the modeling procedure are highlighted with a black edge. {\it Bottom left panel}: Final metallicity offset in Galactic coordinates after the homogenization from ISL + ZSL, $\overline{\Delta} u_{\rm FeH} = \Delta u_{\rm FeH} - F_u$. {\it Bottom right panel} : Distribution of the metallicity offsets $\Delta u_{\rm FeH}$ (gray) and $\overline{\Delta} u_{\rm FeH}$ (colored). The Gaussian distributions that better describe the data are also shown, with their dispersion labeled in the panel. The dotted line marks zero offset.} 
\label{fig:ufeh_2d}
\end{figure*}

\subsubsection{Estimation of the metallicity-dependent stellar locus}\label{sec:ZSLcal}
The stellar locus is known to vary with metallicity \citep[e.g.,][]{yuan15}. Such variation is more prominent at blue optical wavelengths, with the effect in the $u$ band being an order of magnitude larger than in the $g$ band \citep{yuan15}. To illustrate this effect, the median [Fe/H] from LAMOST in the $(u_{\rm ISL}-r)_0$ versus $(g-i)_0$ color-color space is presented in the {\it top panel} of Fig.~\ref{fig:zsl_1d}. At a given $(g-i)_0$ color, redder stars in $(u_{\rm ISL}-r)_0$ have larger metallicities.

As a starting point, we defined the reference stellar locus, noted ${\rm ZSL}_{\rm o}$, from those stars with $-0.25 < {\rm [Fe/H]} < -0.20$ in the color range $0.2 < (g-i)_0 < 1.5$. The metallicity range was chosen to cover the density peak in the LAMOST distribution. From the ZSL$_{\rm o}$ reference, the $(u_{\rm ISL}-r)_0$ color difference to stars of different metallicities was computed, as shown in the {\it bottom panel} of Fig.~\ref{fig:zsl_1d}. We find a $(u_{\rm ISL}-r)_0$ color difference of $-0.20$ mag for ${\rm [Fe/H]} \sim -1.4$ dex stars, and of $+0.23$ mag for ${\rm [Fe/H]} \sim 0.4$ dex stars. The dispersion with respect to the reference locus decreases by a factor of two from $\sigma = 0.092$ mag to $\sigma = 0.047$ mag after accounting for the metallicity dependence. 

As shown by \citet{yuan15}, the metallicity-dependent stellar locus, noted ZSL, is not just a shift from the reference, and the simple modeling described above must be refined. Hence, we estimated the mean $(u_{\rm ISL}-r)_0$ color as a function of [Fe/H] and $(g-i)_0$ with a two-dimensional histogram. The used ranges were $(g-i)_0 \in [0.2,1.5]$ and ${\rm [Fe/H]} \in [-2.1,0.53]$, with 150 bins in each dimension. The ZSL and its difference with respect to the reference locus ZLS$_{\rm o}$ are shown in Fig.~\ref{fig:zsl_2d}, highlighting the shift and the change in curvature of the stellar locus with metallicity. We compare the J-PLUS ZSL with the results from \citet{yuan15} in Sect.~\ref{sec:ZMW}.

\begin{figure*}[t]
\centering
\resizebox{0.49\hsize}{!}{\includegraphics{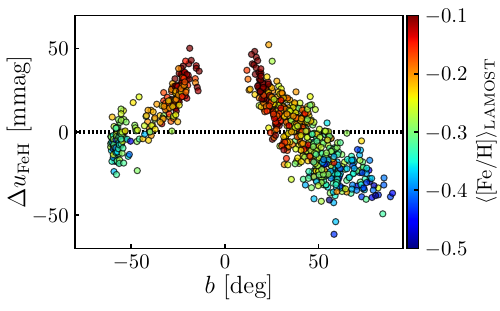}}
\resizebox{0.49\hsize}{!}{\includegraphics{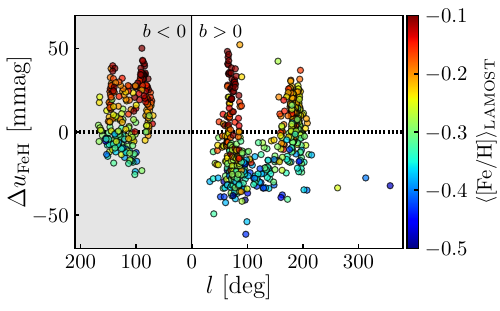}}

\resizebox{0.49\hsize}{!}{\includegraphics{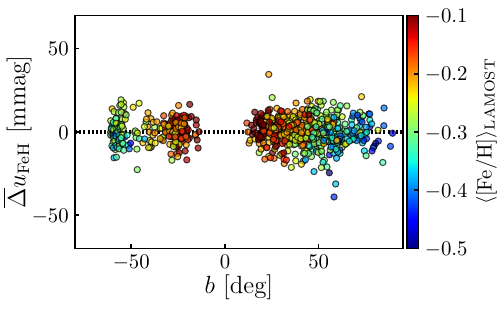}}
\resizebox{0.49\hsize}{!}{\includegraphics{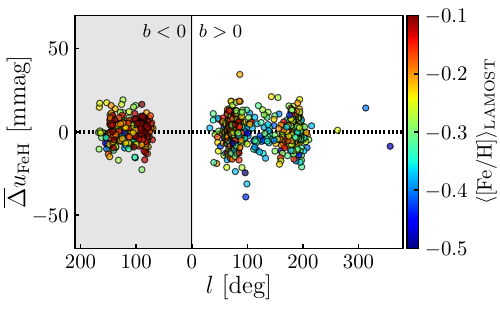}}
\caption{Photometric offset estimated from the metallicity-dependent stellar locus. {\it Top panels}: Initial offset with the homogenization from ISL, $\Delta u_{\rm FeH}$. {\it Bottom panels}: Final offset with the homogenization from ISL + ZS, $\overline{\Delta} u_{\rm FeH} = \Delta u_{\rm FeH} - F_u$. The {\it left panels} show the dependence on Galactic latitude $b$ and the {\it right panels} on Galactic longitude $l$, showing pointings with positive and negative latitudes separately. In all the panels, the color scale shows the median metallicity in the pointing estimated from LAMOST spectra.}
\label{fig:ufeh_1d}
\end{figure*}

\subsubsection{Measurement of the metallicity offset}\label{sec:ZSLoff}
The ZSL estimated in the previous section was used to compute the calibration offset due to metallicity in each J-PLUS DR2 pointing, named $\Delta u_{\rm FeH}$. The expected color for each MS calibration star with LAMOST information is estimated from the ZSL and subtracted to the observed color:
\begin{equation}
\delta u_{\rm FeH} = (u_{\rm ISL}-r)_0 - {\rm ZSL}.
\end{equation}
The distribution of these differences in each pointing was fitted to a Gaussian with median $-\Delta u_{\rm FeH}$, the targeted metallicity offset for the pointing. We assumed that the measured offset is due to the different metallicity, this is, the stellar locus location, of the stars in the pointing with respect to the J-PLUS ISL. In this process, only ZSL bins of more than ten sources and pointings with more than 50 LAMOST stars with $\delta u_{\rm FeH}$ computed were kept.

We illustrate the process using the J-PLUS pointing $p_{\rm id} = 00066$. The dispersion when no metallicity information is included is $\sigma = 0.09$ mag, and a clear dependence with [Fe/H] is present ({\it top panel} in Fig.~\ref{fig:deltafeh}). After accounting for metallicity effects with the ZSL, the dispersion reduces to $\sigma = 0.04$ and the [Fe/H] gradient has disappeared ({\it bottom panel} in Fig.~\ref{fig:deltafeh}). The median of the measured $\delta u_{\rm FeH}$ is $0.038$ mag, and the estimated metallicity offset is therefore $\Delta u_{\rm FeH} = -0.038$ mag.

After applying the above procedure to the $1\,088$ J-PLUS DR2 pointings, we obtained a valid $\Delta u_{\rm FeH}$ for 746 (69\%) of them. In the next section, we study the trends on the derived offsets and detail the treatment of those pointings without a metallicity offset measurement.

\subsubsection{Metallicity offset as a function of the pointing location and iterative process}\label{sec:ZSLmodel}
The metallicity offsets for each J-PLUS pointing with a valid measurement are shown as a function of Galactic coordinates in the {\it top left panel} of Fig.~\ref{fig:ufeh_2d} (2D representation) and in the {\it top panels} of Fig.~\ref{fig:ufeh_1d} (1D representation). We found a systematic trend in the offsets, changing from $\Delta u_{\rm FeH} \sim +0.05$ mag to $\Delta u_{\rm FeH} \sim -0.05$ mag as we move from low to high Galactic latitudes. This trend mirrors the change in the metallicity of the pointings, noted $\langle {\rm [Fe/H]} \rangle_{\rm LAMOST}$ and computed as the median [Fe/H] of the MS calibration stars with LAMOST information, which changes from $-0.1$ dex to $-0.5$ dex (Fig.~\ref{fig:ufeh_1d}).

The dispersion in the distribution of the offsets is $\sigma_{\rm ISL} = 24$ mmag ({\it bottom right panel} in Fig.~\ref{fig:ufeh_2d}), which translates to the observed edge-to-edge difference of $\sim 0.1$ mag. The key feature of the estimated metallicity offsets is their systematic variation, which translates to a systematic shift in the calibration and poses a limitation on the scientific cases that depend on the information from the bluer J-PLUS passbands. As an example, we explore the impact in the estimation of photometric metallicity in Sect.~\ref{sec:ZMW}.

To correct for the metallicity effect, we modeled the variation of $\Delta u_{\rm FeH}$ with Galactic coordinates using a fourth degree polynomial fit: 
\begin{equation}
    F_{u}\,(l,b) = \sum_{m,n = 0}^{4} C_{mn} \times l^m \times b^n,
\end{equation}
where $(l,b)$ are the Galactic longitude and latitude of the J-PLUS pointings, and $C_{mn}$ are the coefficients of the polynomial. This modeling assumes a smooth variation of the calibration offsets (i.e., of the metallicity) along the Galaxy and permits us to assign a metallicity offset to those orphan pointings without a valid measurement. A drawback is that local metallicity variations can still affect the calibration, and in several cases the offsets were extrapolated from the area with available information.

The model $F_u$ was applied as a proxy for the metallicity offset in Eq.~(\ref{eq:zp}). We note that this action changes the photometry of the J-PLUS stars used to compute the ZSL. To ensure self-consistency, we computed an updated version of the ZSL after obtaining the new calibration and iterate the process until convergence. It took four iterations to reach variations lower than 1 mmag in the measured metallicity offsets.

The final $F_u$ model is presented in the {\it top right panel} of Fig.~\ref{fig:ufeh_2d}. The final residuals, noted as $\overline{\Delta} u_{\rm FeH} = \Delta u_{\rm FeH} - F_u$, have a dispersion of $\sigma_{\rm ISL + ZLS} = 7$ mmag, three times smaller than the original ones ({\it bottom right panel} in Fig.~\ref{fig:ufeh_2d}). The improvement is also clear in the lower panels of Figs.~\ref{fig:ufeh_2d} and \ref{fig:ufeh_1d}, where the initial structures are suppressed and no systematic variations with the pointing location remain. This implies that the original systematic error is now statistical, greatly improving the accuracy of the J-PLUS calibration along the surveyed area (Sect.~\ref{sec:accuracy}).
 
As a summary of this section, we estimated and corrected the systematic impact of the varying MW metallicity in the stellar locus calibration. We used the $u$ passband as an illustrative example, and the results for the other J-PLUS passbands are presented in Sect.~\ref{sec:accuracy}.

\section{Error budget and the impact of metallicity on photometric calibration}\label{sect:discussion}
This section is devoted to the error budget analysis and the impact of the metallicity in the J-PLUS DR2 calibration. We study the precision in the photometry in Sect.~\ref{sec:precision} and the accuracy along the surveyed area in Sect.~\ref{sec:accuracy}. The uncertainty in the absolute calibration is discussed in Sect.~\ref{sect:wds}.
\subsection{Precision from overlapping areas}\label{sec:precision}
Adjacent J-PLUS pointings slightly overlap with each other. We measured the precision of the calibration by comparing the photometry of those MS calibration stars independently observed by two pointings. The number of unique pointings pairs in J-PLUS DR2 is $2\,449$. For each pointing pair, we computed the difference between the two calibrated magnitudes of the common stars and estimated the median of the differences to minimize the effect of the individual errors. The distribution of the $2\,449$ median differences was described by a Gaussian function, and the desired precision was obtained as $\sigma/\sqrt{2}$, where $\sigma$ was the measured dispersion of the distribution.

We found that the precision obtained in $\mathcal{X}_{\rm ISL+ZSL}$ magnitudes is similar and replicates the results from J-PLUS DR1 at one mmag level. The results are summarized in Table~\ref{tab:jplus_calib}. The measured precision is $\sim 18$ mmag in $u$, $J0378$, and $J0395$; $\sim 9$ mmag $J0410$ and $J0430$; and $\sim 5$ mmag in $g$, $J0515$, $r$, $J0660$, $i$, $J0861$, and $z$.

We also found that the results with $\mathcal{X}_{\rm ISL}$ magnitudes mimic those in Table~\ref{tab:jplus_calib}. The negligible change with respect to DR1 and after applying the ZSL reflects that metallicity variations along the MW impacts the calibration at scales larger than a few square degrees. This limited local impact is exacerbated when distant pointings are compared, as analyzed in the next section.

\begin{table*} 
\caption{Estimated error budget of the J-PLUS DR2 photometric calibration and final median zero points.} 
\label{tab:jplus_calib}
\centering 
        \begin{tabular}{@{\extracolsep{8pt}}l c c c c c c c c}
        \hline\hline\noalign{\smallskip}
           &  \multicolumn{3}{c}{Precision}  & \multicolumn{4}{c}{Accuracy} \\
        \cline{2-4}\cline{5-8}\noalign{\smallskip}
        Passband    & $\sigma^{\rm pre}_{\rm ISL + ZSL}$   & $\sigma_{\rm WD}$ & $\sigma_{\rm cal}$ & $\sigma^{\rm acc}_{\rm ISL}$ & $s_{\rm ISL}$  & $\sigma^{\rm acc}_{\rm ISL + ZSL}$ & $\sigma^{\rm acc}_{\rm SCR}$&  $\langle {\rm ZP}_{\mathcal{X}}\rangle$\\
                &       [mmag]\tablefootmark{a}   &      [mmag]\tablefootmark{b}          &         [mmag]\tablefootmark{c}        &          [mmag]\tablefootmark{d}  &        [mmag]\tablefootmark{e}           &     [mmag]\tablefootmark{f}   & [mmag]\tablefootmark{g}   & [mag]\\

        \hline\noalign{\smallskip}
        $u$             &  17  &   4  & 18  &  24  &  65  &   7  & 11 & 21.16\\ 
        $J0378$         &  19  &   4  & 20  &  26  &  70  &   8  & 14 & 20.55\\ 
        $J0395$         &  16  &   4  & 17  &  17  &  46  &   6  & 12 & 20.41\\ 
        $J0410$         &  10  &   4  & 12  &  10  &  24  &   4  &  7 & 21.35\\ 
        $J0430$         &   8  &   3  & 10  &   6  &  16  &   3  &  6 & 21.40\\ 
        $g$             &   4  &   2  &  7  &$\cdots$&$\cdots$&$\cdots$&  3 & 23.61\\ 
        $J0515$         &   6  &   2  &  8  &   2  &   4  &   1  &  3 & 21.58\\ 
        $r$             &   4  &$\cdots$&  6  &$\cdots$&$\cdots$&$\cdots$&  3 & 23.65\\ 
        $J0660$         &   5  &   3  &  8  &   1  &   1  &   1  &  3 & 21.12\\ 
        $i$             &   4  &   2  &  7  &$\cdots$&$\cdots$&$\cdots$&  2 & 23.35\\ 
        $J0861$         &   5  &   4  &  8  &   4  &   9  &   2  &  3 & 21.65\\ 
        $z$             &   5  &   3  &  8  &   4  &   9  &   2  &  4 & 22.78\\ 
        \hline 
\end{tabular}
\tablefoot{
\tablefoottext{a} {Instrumental stellar locus (ISL), the plane correction to account for 2D variations along the CCD, and the metallicity-dependent stellar locus (ZSL) were used to homogenize the photometry. The calibration was anchored to PS1 photometry for $gri$ passbands. Precision estimated from duplicated MS stars in overlapping pointings.}\\
\tablefoottext{b} {Uncertainty in the color calibration from the Bayesian analysis of the white dwarf locus (Sect.~\ref{sect:wds}).}\\
\tablefoottext{c} {Final precision in the J-PLUS DR2 flux calibration, $\sigma^2_{\rm cal} = \sigma^2_{\rm ISL+ZSL} + \sigma^2_{\rm WD} + \sigma_r^2$, where $\sigma_r = 5$ mmag (Sect.~\ref{sect:wds})}.\\
\tablefoottext{d} {Dispersion in the metallicity offsets $\Delta \mathcal{X}_{\rm FeH}$ when the ISL was used to homogenize the photometry.}\\
\tablefoottext{e} {Accuracy along the surveyed area estimated from the difference between the median $\Delta \mathcal{X}_{\rm FeH}$ of the 30 pointings with lower and larger absolute latitude when the ISL magnitudes were used (Sect.~\ref{sec:accuracy}).}\\
\tablefoottext{f} {Accuracy along the surveyed area estimated from the  dispersion in $\Delta \mathcal{X}_{\rm FeH}$ when the ISL+ZSL were used to homogenize the photometry.}\\
\tablefoottext{g}{Accuracy estimated from the comparison of the final ISL+ZSL calibration with results from the stellar color regression method (Sect.~\ref{sec:scr}).}
}
\end{table*}

\begin{figure}[t]
\centering
\resizebox{\hsize}{!}{\includegraphics{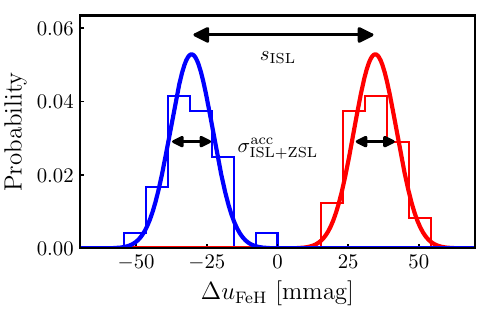}}

\resizebox{\hsize}{!}{\includegraphics{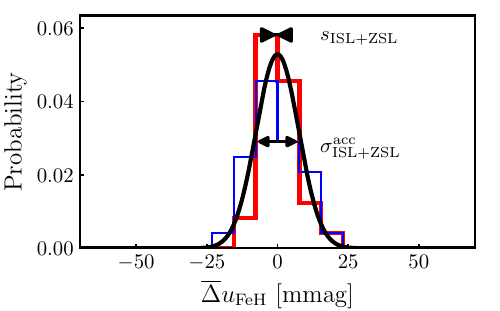}}
\caption{Photometric offset estimated from the metallicity-dependent stellar locus for the 30 J-PLUS DR2 pointings with lower (red histogram) and larger (blue histogram) absolute Galactic latitude. {\it Top panel}: Initial offsets with the homogenization from ISL, $\Delta u_{\rm FeH}$. {\it Bottom panel}: Final offsets with the homogenization from ISL + ZSL, $\overline{\Delta} u_{\rm FeH} = \Delta u_{\rm FeH} - F_u$. The difference between the median offsets of each population is marked as $s_{\rm ISL} = 65$ mmag and $s_{\rm ISL + ZSL} \sim 0$, respectively. The solid lines show Gaussian distributions located at the median offsets and with dispersion $\sigma^{\rm acc}_{\rm ISL + ZSL} = 7$ mmag in all cases, i.e., they are not a fit to the data.}
\label{fig:sisl}
\end{figure}

\begin{figure}[t]
\centering
\resizebox{\hsize}{!}{\includegraphics{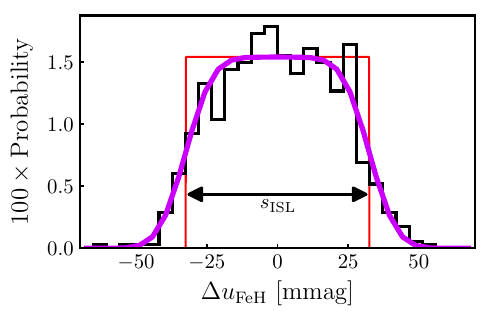}}
\caption{Photometric offset in the $u$ band estimated from the metallicity-dependent stellar locus (open histogram). The red solid line shows a pulse function of width $s_{\rm ISL} = 65$ mmag. The purple solid line is the convolution of the pulse function with a Gaussian function of dispersion $\sigma^{\rm acc}_{\rm ISL + ZSL} = 7$ mmag.}
\label{fig:ufeh_udist}
\end{figure}

\subsection{Accuracy along the surveyed area}\label{sec:accuracy}
The comparison of the photometry in adjacent pointings is not able to provide a measurement of the accuracy of the calibration along the surveyed area. This was a drawback of the analysis performed with J-PLUS DR1 by \citet{clsj19jcal}. As shown in Sect.~\ref{sec:ZSLoff}, the systematic variation of the metallicity along the MW accordingly produces a systematic offset in the photometric solution. The metallicity offsets $\Delta \mathcal{X}_{\rm FeH}$ therefore provide a proxy for the accuracy in the calibration along the J-PLUS DR2 surveyed area.

The dispersion in the metallicity offsets when $\mathcal{X}_{\rm ISL}$ magnitudes were used, noted as $\sigma^{\rm acc}_{\rm ISL}$, is 24 mmag in the $u$ band (Fig.~\ref{fig:ufeh_2d}). This dispersion decreases by a factor of three after accounting for the metallicity effect, $\sigma^{\rm acc}_{\rm ISL + ZSL} = 7$ mmag. However, the systematic nature of the offsets, with a clear smooth variation with Galactic latitude (Fig.~\ref{fig:ufeh_1d}), implies that the relevant measurement of the accuracy in the case of the ISL magnitudes is not the dispersion.

To illustrate this fact, we studied the metallicity offset for the 30 pointing with lower and larger absolute Galactic latitude. Each sample comprises 0.4\% of the total pointings with a metallicity offset measurement. The difference between the median offset of these two samples is $s_{\rm ISL} = 65$ mmag ({\it top panel} in Fig.~\ref{fig:sisl}). Interestingly, we found that the dispersion in the offsets is compatible with $\sigma^{\rm acc}_{\rm ISL + ZSL}$. In other words, the local dispersion before accounting for the metallicity effect is similar to the final dispersion from $\mathcal{X}_{\rm ISL + ZSL}$ magnitudes. By construction, this systematic difference disappeared when $\mathcal{X}_{\rm ISL + ZSL}$ magnitudes were used ({\it bottom panel} in Fig.~\ref{fig:sisl}).

The distribution of $\Delta u_{\rm FeH}$ is therefore not a Gaussian, but the combination of a systematic error function and a random Gaussian component. The systematic error can be expressed with a pulse function:
\begin{equation}
    S\,(x\,|\,s_{\rm ISL}) = H\,(x + 0.5 s_{\rm ISL}) - H\,(x - 0.5 s_{\rm ISL}),
\end{equation}
where $H$ is the Heaviside step function. We interpret this functional form as follows: without the relevant information about the Galactic latitude of the pointing, the value of the metallicity offset should be located between the two measured extremes ($\pm 0.5 s_{\rm ISL}$) with an equal probability. In this context, the accuracy of the ISL calibration is expressed with the variable $s_{\rm ISL}$.

The final distribution of $\Delta u_{\rm FeH}$ was modeled as the convolution of the pulse function $S$ with a Gaussian of dispersion $\sigma^{\rm acc}_{\rm ISL + ZSL}$. We found that this distribution properly describes the observed values (Fig.~\ref{fig:ufeh_udist}), supporting our interpretation. In the case of the $\mathcal{X}_{\rm ISL + ZSL}$ magnitudes, we find $s_{\rm ISL + ZSL} \sim 0$, and only the random component of the error remains.

\begin{figure*}[t]
\centering
\resizebox{0.32\hsize}{!}{\includegraphics{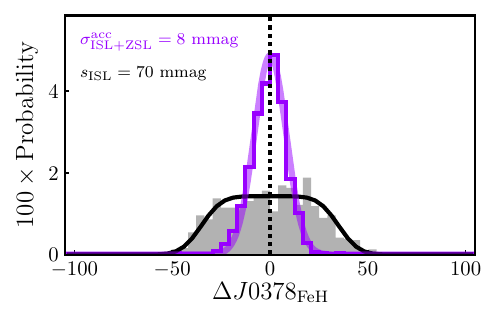}}
\resizebox{0.32\hsize}{!}{\includegraphics{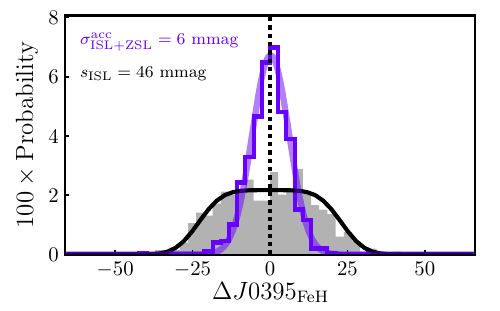}}
\resizebox{0.32\hsize}{!}{\includegraphics{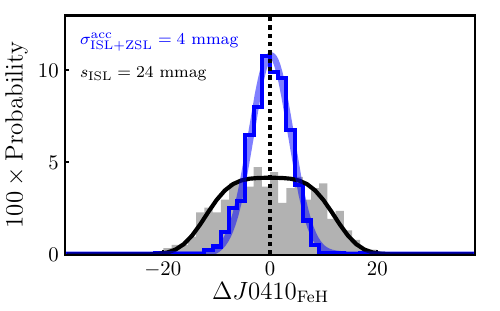}}

\resizebox{0.32\hsize}{!}{\includegraphics{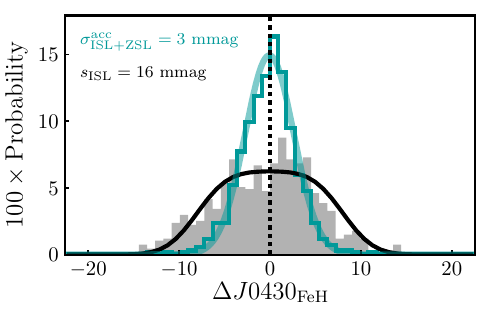}}
\resizebox{0.32\hsize}{!}{\includegraphics{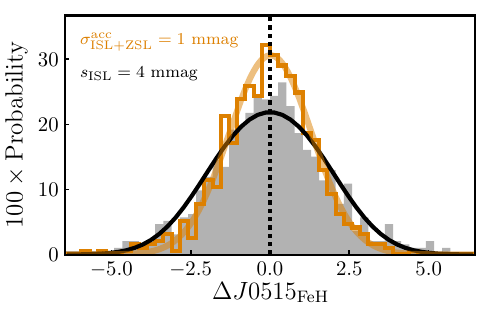}}
\resizebox{0.32\hsize}{!}{\includegraphics{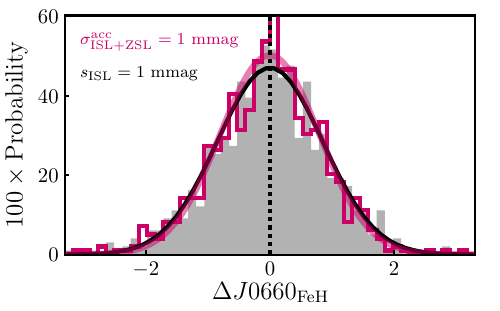}}

\resizebox{0.32\hsize}{!}{\includegraphics{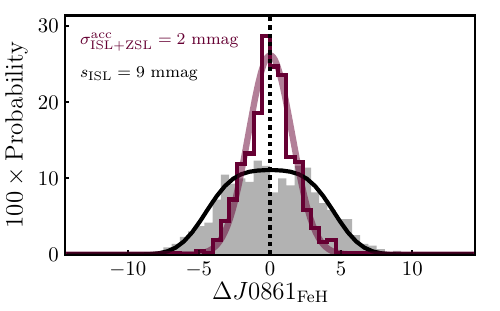}}
\resizebox{0.32\hsize}{!}{\includegraphics{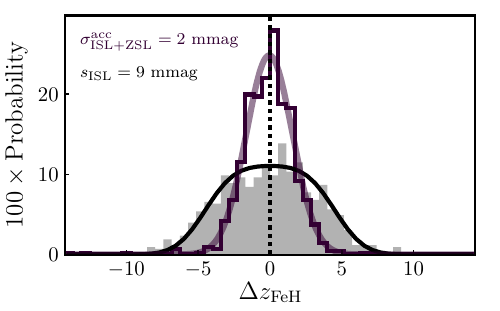}}
\caption{Distribution of the metallicity offsets for J-PLUS DR2. In all the panels, the initial offsets $\Delta \mathcal{X}_{\rm FeH}$ from ISL magnitudes are presented in gray, and the final offsets $\overline{\Delta} \mathcal{X}_{\rm FeH}$ computed with ISL+ZSL magnitudes are in color. From top to bottom and left to right, passbands $J0378$, $J0395$, $J0410$, $J430$, $J0515$, $J0660$, $J0861$, and $z$ are shown. The colored solid line shows the Gaussian distribution that better describe $\overline{\Delta} \mathcal{X}_{\rm FeH}$, with its dispersion $\sigma^{\rm acc}_{\rm ISL + ZSL}$ labeled in each panel. The black solid line depicts the convolution of a pulse function of width $s_{\rm ISL}$ with a Gaussian function of dispersion $\sigma^{\rm acc}_{\rm ISL + ZSL}$.
} 
\label{fig:ufeh_dist}
\end{figure*}

The different accuracy measurements are gathered in Table~\ref{tab:jplus_calib} and shown in Fig.~\ref{fig:ufeh_dist}. We find that the calibration accuracy when metallicity effects are neglected is well above 1\% for the passbands at $\lambda < 4\,500$ \AA\ and accounts for $s_{\rm ISL} \sim 0.07$, $0.07$, $0.05$, $0.03$, and $0.02$ mag in $u$, $J0378$, $J0395$, $J0410$, and $J0430$. The impact is milder in the redder passbands, with $\sim0.01$ mag in $J0861$ and $z$, and negligible in the $J0515$ and $J0660$ passbands.

The numbers above should be representative of the calibration accuracy of J-PLUS DR1, complementing the results presented in \citet{clsj19jcal}. The estimated systematic errors are much larger than the precision errors, limiting the J-PLUS scientific outcome when using the DR1 calibration. This is illustrated in Sect.~\ref{sec:ZMW}.

The implementation of the metallicity-dependent stellar locus has greatly improved the accuracy in the J-PLUS calibration. It has not only decreased the dispersion in the final metallicity offsets by a factor of two-three (Fig.~\ref{fig:ufeh_dist}), but also removed the main systematic error. The final uncertainty estimated for J-PLUS DR2 is summarized in Table~\ref{tab:jplus_calib} and is at the 1\% level or below. The improvement in the bluer passbands is roughly a factor ten, decreasing from $s_{\rm ISL} \sim 70 - 20$ mmag to $\sigma_{\rm ISL+ZSL}^{\rm acc} \sim 8 - 3$ mmag.

We conclude that the implementation of the metallicity-dependent stellar locus has improved the accuracy of the J-PLUS DR2 calibration to 1\% level and has minimized the systematic in the photometric solution along the surveyed area.

\subsection{Absolute color calibration from the white dwarf locus}\label{sect:wds}
The stellar locus steps, both ISL and ZSL, were devoted to homogenizing J-PLUS photometry in those passbands not anchored to PS1. The absolute color calibration was performed with the white dwarf locus. Thanks to the large area observed ($2\,176$ deg$^2$) and the already homogenized photometry, a set of $639$ high-quality white dwarfs were retrieved from the {\it Gaia} absolute magnitude versus color diagram. We performed a joint Bayesian analysis of the eleven $(\mathcal{X}_{\rm ISL+ZSL} - r)_0$ versus $(g-i)_0$ color-color diagrams to estimate the offsets $\Delta \mathcal{X}_{\rm WD}$ that translate instrumental magnitudes to calibrated magnitudes on top of the atmosphere. We summarize the obtained values in Table~\ref{tab:wd_model} for reference and provide a detailed technical description of the process in Appendix~\ref{app:wdlocus}. The typical uncertainty in these offsets is at the $4$ mmag level, as presented also in Table~\ref{tab:jplus_calib}.

In addition to the offsets, the Bayesian modeling provides the intrinsic dispersion in the WD locus (Table~\ref{tab:wd_model}) and two physical parameters of the WD population: the fraction of DAs and the median gravity. We found a DA fraction of $f_{\rm DA} = 0.83 \pm 0.01$ and a median $\log {\rm g} = 7.97 \pm 0.04$. Both values are consistent with J-PLUS DR1 results in \citet{clsj19jcal}, and the median surface gravity agrees with the literature \citep[e.g.,][and references therein]{jimenezesteban18,gentilefusillo19,tremblay19,bergeron19}. We refer the reader to \citet{clsj19jcal} and Appendix~\ref{app:wdlocus} for a detailed description of the Bayesian modeling and the assumptions in the white dwarf locus step.

A relevant change with respect to J-PLUS DR1 is on the $g$ and $i$ passbands' offsets. We obtained $\Delta g_{\rm WD} = 1$ mmag and $\Delta i_{\rm WD} = -1$ mmag, while the values in DR1 were $\Delta g_{\rm WD} = -3$ mmag and $\Delta i_{\rm WD} = 4$ mmag, respectively. We attribute the better agreement between J-PLUS and PS1 photometric systems to the change in the assumed color excess and the extinction coefficients (Sect.~\ref{sec:ext}).

We set the calibration uncertainty in the reference $r$ band to $\sigma_r = 5$ mmag following the results in DR1. This uncertainty is added to the precision in the homogenization and the white dwarf locus offsets to provide the absolute flux uncertainty in J-PLUS DR2 (Table~\ref{tab:jplus_calib}). The final precision is comparable to DR1 and the new calibration considerably improves the accuracy of our photometry.

As a final test, we compared the J-PLUS magnitudes with the synthetic photometry of the spectroscopic standard white dwarf GD 153. We summarize the details of the comparison in Appendix~\ref{app:gd153}. We found that the agreement is at the $3$\% level in $u$; at the $2$\% level in $J0378, J0395, J0410$, and $J430$; and at the $1$\% level in $g$, $J0515$, $r$, $J0660$, $i$, $J0861$, and $z$. These differences were set as an upper limit for the absolute flux calibration accuracy of the J-PLUS DR2 photometry.

\begin{table} 
\caption{Estimated offsets to transport the ISL+ZSL photometry outside the atmosphere.} 
\label{tab:wd_model}
\centering 
        \begin{tabular}{l c c}
        \hline\hline\rule{0pt}{3ex} 
        Passband $(\mathcal{X})$   &   $\Delta \mathcal{X}_{\rm WD}$    &   $\sigma_{\rm int}$ \\\rule{0pt}{2ex} 
                &   [mag]                & [mag]   \\
        \hline\rule{0pt}{2ex}
        $u$             &$-3.881 \pm 0.004$       &$0.030 \pm 0.005$          \\ 
        $J0378$         &$-4.497 \pm 0.004$       &$0.028 \pm 0.005$          \\ 
        $J0395$         &$-4.616 \pm 0.004$       &$0.028 \pm 0.005$          \\ 
        $J0410$         &$-3.662 \pm 0.004$       &$0.008 \pm 0.005$          \\ 
        $J0430$         &$-3.603 \pm 0.003$       &$0.007 \pm 0.004$          \\ 
        $g$             &$\ \ \ 0.001 \pm 0.002$  &$0.004 \pm 0.002$          \\ 
        $J0515$         &$-3.438 \pm 0.002$       &$0.008 \pm 0.003$          \\ 
        $r$             &$\cdots$                 &$\cdots$                   \\ 
        $J0660$         &$-3.901 \pm 0.003$       &$0.019 \pm 0.003$          \\ 
        $i$             &$-0.001 \pm 0.002$       &$0.003 \pm 0.002$          \\ 
        $J0861$         &$-3.371 \pm 0.004$       &$0.013 \pm 0.006$          \\ 
        $z$             &$-2.250 \pm 0.003$       &$0.011 \pm 0.004$          \\ 
        \hline
\end{tabular}
\end{table}

\begin{figure*}[t]
\centering
\resizebox{0.49\hsize}{!}{\includegraphics{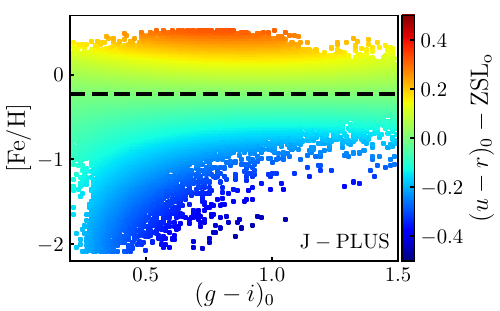}}
\resizebox{0.49\hsize}{!}{\includegraphics{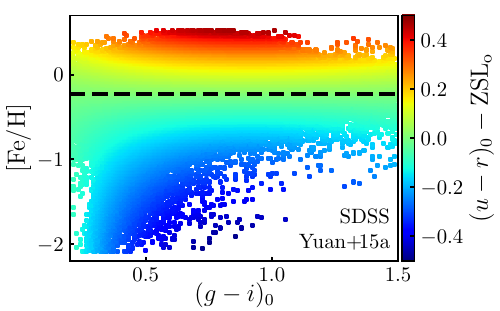}}
\caption{Modeled $(u - r)_0$ color difference with respect to the reference locus, ZSL$_{\rm o}$, as a function of $(g-i)_0$ and [Fe/H]. The median metallicity of the reference locus is marked with the black dashed line. {\it Left panel}: Estimation from J-PLUS DR2 final photometry. {\it Right panel}: Estimation from \citet{yuan15} using SDSS photometry. 
}
\label{fig:zsl_final}
\end{figure*}

\subsection{Impact of the assumed color excess in the calibration}\label{sect:ebvtest}
We compared the final zero points obtained with the stellar and white dwarf loci against the zero points obtained by direct comparison with the PS1 $z$ passband. The difference is well described by a Gaussian with median $\mu = -2$ mmag and a dispersion of $\sigma = 5$ mmag. This result reinforces the calibration procedure and was used to discriminate the best extinction model. 

We repeated the full calibration process assuming the color excess at infinity from \citet{planck_dust} and with the 3D dust maps provided by \texttt{Bayestar17}\footnote{\url{http://argonaut.skymaps.info/}} \citep{bayestar17}, based on Pan-STARRS stellar colors. We found that the best consistency with the PS1 $z$-band photometry is reached with the estimation based on \citet{sfd98}. In all the cases, the systematic offsets due to metallicity were present.

Interestingly, the application of the metallicity-dependent stellar locus to the $\mathcal{X}_{\rm ISL}$ magnitudes worsens the comparison between J-PLUS and PS1 in the \texttt{Bayestar17} case, going for $\sigma = 5$ mmag to $\sigma = 7$ mmag. The opposite happened in the Planck and \citet{sfd98} cases, improving from $\sigma = 6$ mmag to $\sigma = 5$ mmag. The differences are subtle, but measurable. The extinction maps from \citet{sfd98} and \citet{planck_dust} are not related to the photometry used in the calibration, therefore providing an independent extinction frame for the homogenization process.

\subsection{Comparison with the SCR method}\label{sec:scr}
As already pointed out in Sect.~\ref{sec:intro}, the stellar color regression (SCR; \citealt{scr,huang21}) method deals with the different stellar properties in a consistent way, providing an alternative homogenization process for calibration. Using LAMOST DR5 as reference, the SCR method was applied to J-PLUS DR2. 

We found that the comparison between the ISL+ZSL and the SCR zero points follows a Gaussian distribution with dispersion $\sigma^{\rm acc}_{\rm SCR}$,  as reported in Table~\ref{tab:jplus_calib}. The dispersion is $\sim12$ mmag in the $u$, $J0378$, and $J0395$ filters, $\sim6$ mmag in $J0410$ and $J0430$, and $\sim 3$ mmag in the rest of the J-PLUS passbands. The origin of this dispersion is related to the different treatment of the interstellar extinction, our functional approach to the impact of the metallicity offset, and the inherent statistical dispersion of each method. 

A detailed application and analysis of the SCR calibration for J-PLUS DR2 is beyond the scope of the present paper and will be presented in a forthcoming work. The comparison with the independent SCR method provided an extra measurement for the accuracy in the photometry, which we set at a percentage level for passbands bluer than $\lambda \sim 4\,500$ \AA.

\begin{figure*}[t]
\centering
\resizebox{0.49\hsize}{!}{\includegraphics{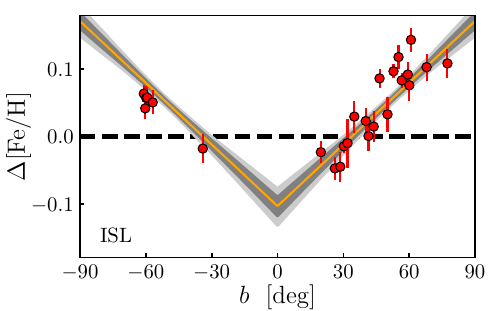}}
\resizebox{0.49\hsize}{!}{\includegraphics{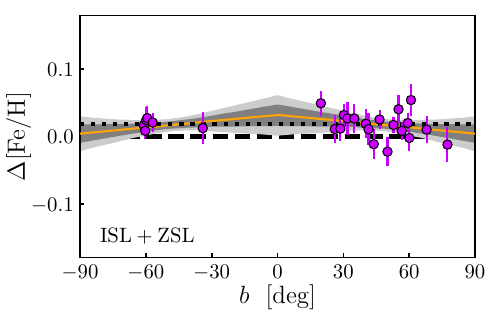}}
\caption{Metallicity difference between J-PLUS photometric values and APOGEE spectroscopic values, $\Delta {\rm [Fe/H]}$, as a function of Galactic latitude $b$. {\it Left panel}: Using $\mathcal{X}_{\rm ISL}$ photometry. {\it Right panel}: Using $\mathcal{X}_{\rm ISL+ZSL}$ photometry. The solid lines in both panels show the best linear fitting to the data, with the gray areas depicting the 68\% and 95\% confidence intervals. The dashed lines mark zero difference. The dotted line in the {\it right panel} shows a difference of 0.02 dex.
} 
\label{fig:feh_apogee}
\end{figure*}

\begin{figure}[t]
\centering
\resizebox{\hsize}{!}{\includegraphics{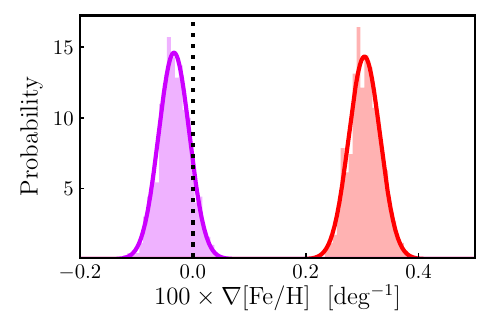}}
\caption{Distribution in the best linear-fitting slope of the metallicity difference versus Galactic latitude estimated from $\mathcal{X}_{\rm ISL}$ (red) and $\mathcal{X}_{\rm ISL+ZSL}$ (purple) photometry. The dotted line marks a zero slope.} 
\label{fig:slope_apogee}
\end{figure}

\subsection{Photometric metallicity from J-PLUS data}\label{sec:ZMW}
In this section, we highlight the impact of the improved calibration in the estimation of the photometric metallicity from J-PLUS DR2 data. As in other sections, we use the $u$ band as an example, but similar results are obtained with J-PLUS passbands $J0378$ and $J0395$, which are the most sensitive to metallicity.

We started by computing the final ZSL in the $(u-r)_0$ versus $(g-i)_0$ space, as in Sect.~\ref{sec:ZSL}, but using the final J-PLUS DR2 calibrated magnitudes. Following \citet{yuan15}, we modeled the $(u-r)_0$ locus with a fourth-degree polynomial in $(g-i)_0$ and [Fe/H]. The resulting model in those bins with data was normalized to the locus at [Fe/H] = -0.225 dex, as shown in the {\it left panel} of Fig.~\ref{fig:zsl_final}. The curvature in the locus is evident.

We compared the J-PLUS ZSL with the results from \citet{yuan15} using SDSS photometry. They provide the metallicity-dependent stellar loci $(u-g)_0$ and $(g-r)_0$ as a function of $(g-i)_0$ and [Fe/H]. We combined both loci to obtain the $(u-r)_0$ variation and normalized again to the locus at [Fe/H]$ = -0.225$ dex. The result is presented in the {\it right panel} of Fig.~\ref{fig:zsl_final}. We found a close agreement between both studies, which present similar structures. The discrepancies, at the 0.04 mag level, are expected because of the different photometric systems used (J-PLUS versus SDSS).

After checking our final ZSL with the results in \citet{yuan15}, we aim to test the impact of the calibration in the photometric metallicities estimated from J-PLUS DR2. We decided to compute the J-PLUS photometric metallicities using the simplest offset model, relating the $(u-r)_0$ color distance to the reference locus at [Fe/H]$ = -0.225$ dex with a metallicity measurement. We used $144\,375$ stars in common with LAMOST and with $0.1 \leq (g-i)_0 \leq 1.6$ to map the relation between color and metallicity. The comparison between J-PLUS and LAMOST metallicities had a dispersion of $\sigma = 0.14$ dex. We stress that the goal of this section is just to illustrate the net improvement of the photometric calibration. We expect to obtain better metallicity estimates from the whole twelve-band J-PLUS photometry \citep[e.g.,][]{whitten19}.

Because LAMOST metallicities were used in both the calibration and the estimation of the photometric metallicity, we ensured an independent test by comparing J-PLUS metallicities with the spectroscopic values from the Apache Point Observatory Galactic Evolution Experiment (APOGEE, \citealt{apogee_dr16}) data in the SDSS DR16\footnote{\url{https://www.sdss.org/dr16/irspec/dr_synopsis/}}. The available data contain high-resolution ($R\sim22\,500$) near-infrared ($15\,140 - 16\,940$ \AA) spectra for about $430\,000$ stars covering both the northern and southern skies, from which radial velocities, stellar parameters, and chemical abundances of 20 species are determined. 

We cross-matched the MS calibration stars with the APOGEE sample using 1 arcsec radius. A total flag equal to zero in APOGEE information and a J-PLUS color $0.1 \leq (g-i)_0 \leq 1.6$ was imposed. This yielded $2\,177$ common stars. The difference between the J-PLUS and APOGEE values was defined as
\begin{equation}
\Delta {\rm [Fe/H]} = {\rm [Fe/H]}_{\rm J-PLUS} - {\rm [Fe/H]}_{\rm APOGEE}.
\end{equation}
The star-by-star difference defined a Gaussian with median of $\mu = 0.03$ dex and a dispersion of $\sigma = 0.13$ dex.

To explore the possible systematic trend of $\Delta {\rm [Fe/H]}$ with Galactic latitude, we computed the median metallicity difference with respect to APOGEE using 25 bins of variable size to ensure $\sim 90$ sources per bin. The uncertainties where estimated by bootstrapping. The results using $\mathcal{X}_{\rm ISL}$ magnitudes and the final calibration are presented in Fig.~\ref{fig:feh_apogee}. We found that the metallicity differences are roughly flat with the final J-PLUS DR2 calibration, as desired, presenting a slight bias of 0.02 dex. However, neglecting the ZSL step in the calibration produces a clear trend with Galactic latitude: the estimated $\Delta {\rm [Fe/H]}$ changes from $-0.02$ dex at $|b| \sim 30$ deg to $+0.10$ dex at $|b| \sim 80$ deg. We performed a linear fit to the data, using $|b|$ as an independent variable, and we present the distribution of the slope $\nabla [\rm Fe/H]$, with [deg$^{-1}$] units, in Fig.~\ref{fig:slope_apogee}. The slope for the final calibration is $100 \times \nabla {\rm [Fe/H]} = -0.03 \pm 0.03$, while neglecting the ZSL step provides $100 \times \nabla {\rm [Fe/H]} = 0.30 \pm 0.03$. The slope is compatible with zero, as desired, by including the impact of metallicity in the stellar locus position, while the slope is positive at $10\sigma$ level when the metallicity effects are not accounted for.
We conclude that the improved photometric calibration of J-PLUS DR2 yields a reliable twelve-bands photometric catalog for an important fraction of the northern sky.

\section{Summary and conclusions}\label{sec:summary}
We explored the impact of metallicity on the photometric calibration of J-PLUS DR2, based on the stellar locus technique, and update the error budget in the calibration.
Using the metallicity information from LAMOST, we find that the J-PLUS passbands bluer than $4\,500$ \AA{} are strongly affected by the MW metallicity gradient in Galactic latitude, which breaks the assumption of a homogeneous dust de-reddened stellar locus across the sky. The peak-to-peak variation amounts to $0.07$, $0.07$, $0.05$, $0.03$, and $0.02$ mag in $u$, $J0378$, $J0395$, $J0410$, and $J0430$, respectively. The variation is $\sim 0.01$ mag in $J0861$ and $z$, while negligible in $J0515$ and $J0660$. This effect is systematic and smooth along the surveyed area. We modeled the metallicity-dependent offset in the stellar locus in those areas in common with LAMOST to improve the photometric calibration in the complete J-PLUS DR2 dataset. The accuracy of the calibration in the surveyed area is expected to be at a percentage level for the bluer J-PLUS passbands and at a sub-percentage level in the rest of the filters after including the metallicity information in the process. The published J-PLUS DR2 photometry already includes the metallicity-dependent stellar locus step in the calibration procedure.

The precision in the calibration, measured from repeated sources in the overlapping areas between pointings and including absolute color and flux scale uncertainties, is $\sim 18$ mmag in $u$, $J0378$, and $J0395$; $\sim 11$ mmag in $J0410$ and $J0430$; and $\sim 8$ mmag in $g$, $J0515$, $r$, $J0660$, $i$, $J0861$, and $z$. These values are similar to those derived in \citet{clsj19jcal} with J-PLUS DR1 data, reflecting that the metallicity impacts the calibration at scales above a few square degrees.

Our analysis highlights the expected impact of metallicity on the stellar locus technique at $\lambda \lesssim 4\,500$ \AA{} (see \citealt{high09,yuan15}), producing systematic offsets and impacting the physical properties derived for stars and galaxies. Large-area surveys with blue optical passbands must evaluate the impact of metallicity in the use of the stellar locus to homogenize their photometry, and techniques based on large overlapping areas or methods that accounts for the variety of stars' physical properties (e.g., SCR or ISL+ZSL) should be favored.

The main limitation of the presented methodology is its dependence on external datasets. We relied on \cite{sfd98} to derive the extinction from the interstellar medium, {\it Gaia} to define a sample of main sequence stars, PS1 to set the calibration of the $gri$ broad bands, and LAMOST to access the metallicity of the calibration stars. The possible systematics from each work will be inherited by the J-PLUS photometry. Because of this and to ensure consistency with the calibration process, we recommend assuming \cite{sfd98} maps and the extinction coefficients in Table~\ref{tab:JPLUS_filters} when using J-PLUS DR2 photometry.

Regarding the technical goal of J-PLUS, which is to ensure the photometric calibration of J-PAS, the findings here are of great importance when defining the J-PAS calibration strategy. The current roadmap for J-PAS calibration has three steps: (1) homogenization using half-CCD overlapping areas thanks to a large dithering pattern between the four exposures per filter, which will allow us to derive a consistent photometric solution along the surveyed area and to trace 2D variations along the focal plane by comparing four measurements of the same source; (2) absolute color calibration using the white dwarf locus, whereby the properties of the locus, with two populations and curved profiles, will permit the color calibration without using external photometric data; and (3) absolute calibration by anchoring the J-PAS reference broad band to Pan-STARRS. In this case, only one offset will be needed to translate the already homogeneous photometry outside the atmosphere. The calibration against {\it Gaia} is also a possibility, but with J-PAS photometry being independent of {\it Gaia} spectro-photometry it will be possible to test systematic effects in both surveys.

\begin{acknowledgements}
We dedicate this paper to the memory of our six IAC colleagues and friends who met with a fatal accident in Piedra de los Cochinos, Tenerife, in February 2007, with  special thanks to Maurizio Panniello, whose teachings of \texttt{python} were so important for this paper.

We thank the anonymous referee for useful comments and suggestions.

We thank the relevant discussions with the J-PLUS collaboration members. 

Based on observations made with the JAST80 telescope at the Observatorio Astrof\'{\i}sico de Javalambre (OAJ), in Teruel, owned, managed, and operated by the Centro de Estudios de F\'{\i}sica del  Cosmos de Arag\'on. We acknowledge the OAJ Data Processing and Archiving Unit (UPAD) for reducing the OAJ data used in this work.

Funding for the J-PLUS Project has been provided by the Governments of Spain and Arag\'on through the Fondo de Inversiones de Teruel; the Aragonese Government through the Reseach Groups E96, E103, and E16\_17R; the Spanish Ministry of Science, Innovation and Universities (MCIU/AEI/FEDER, UE) with grants PGC2018-097585-B-C21 and PGC2018-097585-B-C22; the Spanish Ministry of Economy and Competitiveness (MINECO) under AYA2015-66211-C2-1-P, AYA2015-66211-C2-2, AYA2012-30789, and ICTS-2009-14; and European FEDER funding (FCDD10-4E-867, FCDD13-4E-2685). The Brazilian agencies FINEP, FAPESP, and the National Observatory of Brazil have also contributed to this project.

J.~V. acknowledges the technical members of the UPAD for their invaluable work: Juan Castillo, Tamara Civera, Javier Hern\'andez, \'Angel L\'opez, Alberto Moreno, and David Muniesa.

E.~J.~A acknowledges financial support from PGC2018-095049-B-C21 (MCIU/AEI/FEDER, UE) and SEV-2017-0709.

A.~A.~C. acknowledges support from the Universidad de Alicante (contract UATALENTO18-02).

F.~J.~E. acknowledges financial support from the Spanish MINECO/FEDER through the grant AYA2017-84089 and MDM-2017-0737 at Centro de Astrobiología (CSIC-INTA), Unidad de Excelencia María de Maeztu, and from the European Union’s Horizon 2020 research and innovation programme under Grant Agreement no. 824064 through the ESCAPE - The European Science Cluster of Astronomy \& Particle Physics ESFRI Research Infrastructures project.

The work of V.~M.~P. is supported by NOIRLab, which is managed by the Association of Universities for Research in Astronomy (AURA) under a cooperative agreement with the National Science Foundation.

E.~T. acknowledges support by ETAg grant PRG1006 and by EU through the ERDF CoE grant TK133.

L.~S.~J. acknowledges support from Brazilian agencies FAPESP (2019/10923-5) and CNPq (304819/201794).

Guoshoujing Telescope (the Large Sky Area Multi-Object Fiber Spectroscopic Telescope LAMOST) is a National Major Scientific Project built by the Chinese Academy of Sciences. Funding for the project has been provided by the National Development and Reform Commission. LAMOST is operated and managed by the National Astronomical Observatories, Chinese Academy of Sciences.

The Pan-STARRS1 Surveys (PS1) and the PS1 public science archive have been made possible through contributions by the Institute for Astronomy, the University of Hawaii, the Pan-STARRS Project Office, the Max-Planck Society and its participating institutes, the Max Planck Institute for Astronomy, Heidelberg, and the Max Planck Institute for Extraterrestrial Physics, Garching, The Johns Hopkins University, Durham University, the University of Edinburgh, the Queen's University Belfast, the Harvard-Smithsonian Center for Astrophysics, the Las Cumbres Observatory Global Telescope Network Incorporated, the National Central University of Taiwan, the Space Telescope Science Institute, the National Aeronautics and Space Administration under Grant No. NNX08AR22G issued through the Planetary Science Division of the NASA Science Mission Directorate, the National Science Foundation Grant No. AST-1238877, the University of Maryland, Eotvos Lorand University (ELTE), the Los Alamos National Laboratory, and the Gordon and Betty Moore Foundation.

This work has made use of data from the European Space Agency (ESA) mission
{\it Gaia} (\url{https://www.cosmos.esa.int/gaia}), processed by the {\it Gaia} Data Processing and Analysis Consortium (DPAC, \url{https://www.cosmos.esa.int/web/gaia/dpac/consortium}). Funding for the DPAC has been provided by national institutions, in particular the institutions participating in the {\it Gaia} Multilateral Agreement.

Funding for SDSS-III has been provided by the Alfred P. Sloan Foundation, the Participating Institutions, the National Science Foundation, and the U.S. Department of Energy Office of Science. The SDSS-III web site is \url{http://www.sdss3.org/}.

SDSS-III is managed by the Astrophysical Research Consortium for the Participating Institutions of the SDSS-III Collaboration including the University of Arizona, the Brazilian Participation Group, Brookhaven National Laboratory, Carnegie Mellon University, University of Florida, the French Participation Group, the German Participation Group, Harvard University, the Instituto de Astrofisica de Canarias, the Michigan State/Notre Dame/JINA Participation Group, Johns Hopkins University, Lawrence Berkeley National Laboratory, Max Planck Institute for Astrophysics, Max Planck Institute for Extraterrestrial Physics, New Mexico State University, New York University, Ohio State University, Pennsylvania State University, University of Portsmouth, Princeton University, the Spanish Participation Group, University of Tokyo, University of Utah, Vanderbilt University, University of Virginia, University of Washington, and Yale University.

Funding for the Sloan Digital Sky 
Survey IV has been provided by the 
Alfred P. Sloan Foundation, the U.S. 
Department of Energy Office of 
Science, and the Participating 
Institutions. 

SDSS-IV acknowledges support and 
resources from the Center for High 
Performance Computing  at the 
University of Utah. The SDSS 
website is \url{www.sdss.org}.

SDSS-IV is managed by the 
Astrophysical Research Consortium 
for the Participating Institutions 
of the SDSS Collaboration including 
the Brazilian Participation Group, 
the Carnegie Institution for Science, 
Carnegie Mellon University, Center for 
Astrophysics | Harvard \& 
Smithsonian, the Chilean Participation 
Group, the French Participation Group, 
Instituto de Astrof\'isica de 
Canarias, The Johns Hopkins 
University, Kavli Institute for the 
Physics and Mathematics of the 
Universe (IPMU) / University of 
Tokyo, the Korean Participation Group, 
Lawrence Berkeley National Laboratory, 
Leibniz Institut f\"ur Astrophysik 
Potsdam (AIP),  Max-Planck-Institut 
f\"ur Astronomie (MPIA Heidelberg), 
Max-Planck-Institut f\"ur 
Astrophysik (MPA Garching), 
Max-Planck-Institut f\"ur 
Extraterrestrische Physik (MPE), 
National Astronomical Observatories of 
China, New Mexico State University, 
New York University, University of 
Notre Dame, Observat\'ario 
Nacional / MCTI, The Ohio State 
University, Pennsylvania State 
University, Shanghai 
Astronomical Observatory, United 
Kingdom Participation Group, 
Universidad Nacional Aut\'onoma 
de M\'exico, University of Arizona, 
University of Colorado Boulder, 
University of Oxford, University of 
Portsmouth, University of Utah, 
University of Virginia, University 
of Washington, University of 
Wisconsin, Vanderbilt University, 
and Yale University.

This research made use of \texttt{Astropy}, a community-developed core \texttt{Python} package for Astronomy \citep{astropy}, and \texttt{Matplotlib}, a 2D graphics package used for \texttt{Python} for publication-quality image generation across user interfaces and operating systems \citep{pylab}.
\end{acknowledgements}

\bibliographystyle{aa}
\bibliography{biblio}

\begin{appendix}

\section{Technical details about the absolute color calibration}\label{app:wdlocus}

\begin{figure}[t]
\centering
\resizebox{\hsize}{!}{\includegraphics{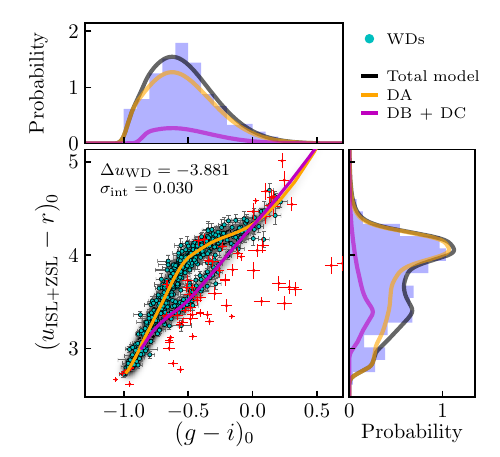}}
\caption{Dust de-reddened $(u_{\rm ISL+ZSL}-r)_0$ versus $(g-i)_0$ color-color diagram of the 639 high-quality white dwarfs in J-PLUS DR2 (clean sample, cyan dots; outliers, red dots). The solid lines show the theoretical locus for DA (orange) and DB+DC WDs (magenta). The gray scale shows the most probable model that describes the observations. The blue probability distributions above and to the right show the $(g-i)_0$ and $(u_{\rm ISL+ZSL} - r)_0$ projections of the data, respectively. The projections of the total, DA, and DB+DC models are represented with the black, orange, and magenta lines. The model in all the J-PLUS color-color diagrams shares the parameters $\mu = -0.834$, $s = 0.385$, $\alpha = 1.50$, $f_{\rm DA} = 0.835$, $\log {\rm g} = 7.97$, and $\Delta \mathcal{C}_{1} = 0.002$ (see text for details). The values of the filter-dependent parameters $\sigma_{\rm int}$ and $\Delta \mathcal{X}_{\rm WD}$ are labeled in the panel.}
\label{fig:wdlocus_1}
\end{figure}

\begin{figure*}[t]
\centering
\resizebox{0.49\hsize}{!}{\includegraphics{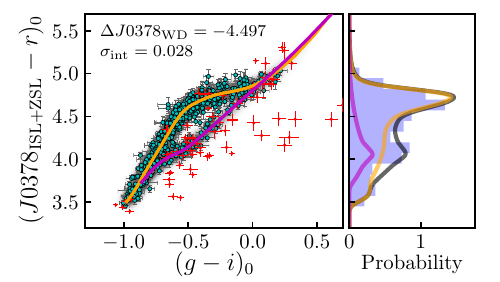}}
\resizebox{0.49\hsize}{!}{\includegraphics{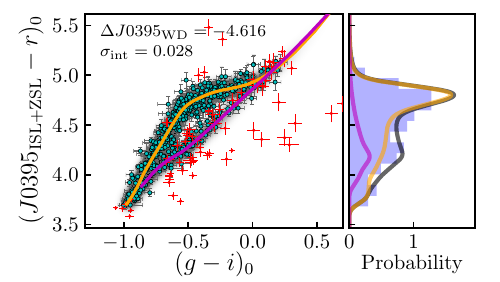}}

\resizebox{0.49\hsize}{!}{\includegraphics{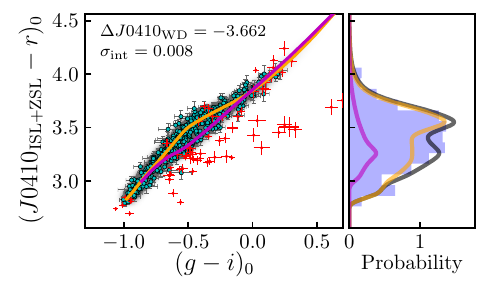}}
\resizebox{0.49\hsize}{!}{\includegraphics{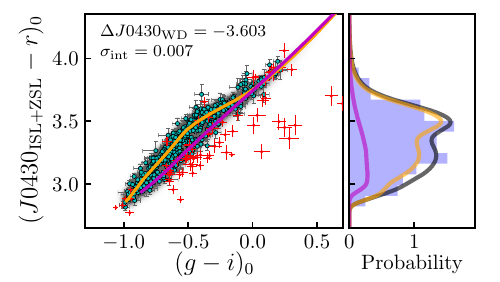}}

\resizebox{0.49\hsize}{!}{\includegraphics{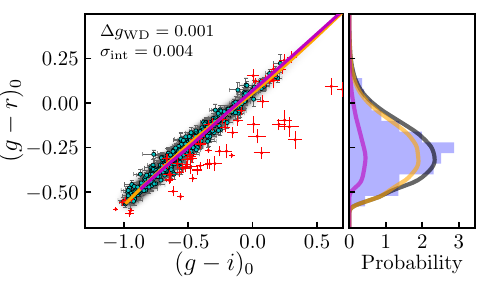}}
\resizebox{0.49\hsize}{!}{\includegraphics{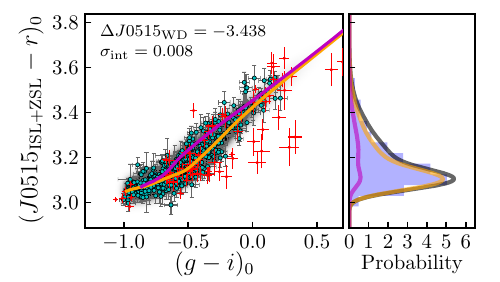}}
\caption{Similar to Fig.~\ref{fig:wdlocus_1}, but for $\mathcal{X}$ = $J0378$, $J0395$, $J0410$, $J0430$, $g$, and $J0515$ passbands. We omit the $(g-i)_0$ projection because it is shared by all the panels.}
\label{fig:wdlocus_2}
\end{figure*}

\begin{figure*}[t]
\centering
\resizebox{0.49\hsize}{!}{\includegraphics{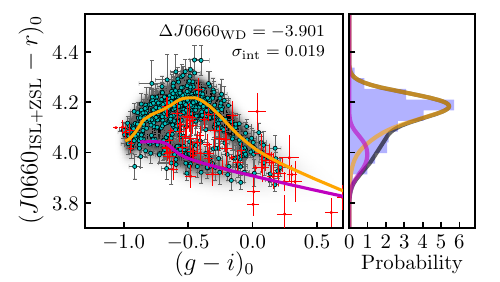}}
\resizebox{0.49\hsize}{!}{\includegraphics{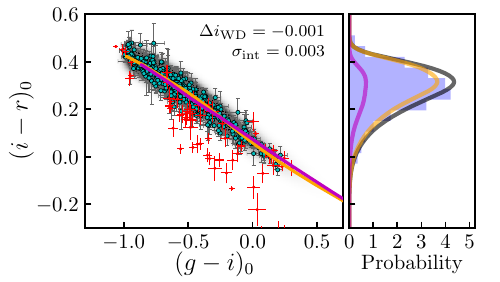}}

\resizebox{0.49\hsize}{!}{\includegraphics{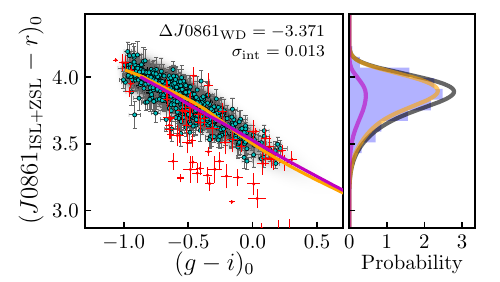}}
\resizebox{0.49\hsize}{!}{\includegraphics{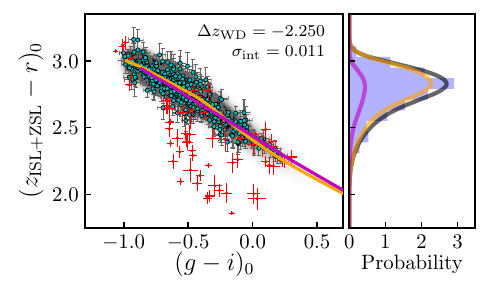}}
\caption{Similar to Fig.~\ref{fig:wdlocus_1}, but for $\mathcal{X}$ = $J0660$, $i$, $J0861$, and $z$ passbands. We omit the $(g-i)_0$ projection because it is shared by all the panels.}
\label{fig:wdlocus_3}
\end{figure*}

The principal novelty presented in \citet{clsj19jcal} is the implementation of the white dwarf (WD) locus for absolute color calibration. Here, we provide the relevant technical details for completeness, and the reader is referred to \citet{clsj19jcal} for a detailed description.

The properties of WDs make them excellent standard sources for calibration \citep{holberg06}. The model atmospheres of WDs can be specified at the $\sim1$\% flux level with an effective temperature ($T_{\rm eff}$) and a surface gravity ($\log {\rm g}$). These parameters can be accurately estimated via a spectroscopic analysis of the Balmer line profiles, providing a reference flux model for calibration. They are also mostly photometrically stable and statistically present lower levels of interstellar reddening than main sequence stars. Because of these properties, a significant observational and theoretical effort is still underway to provide the best possible WD network to ensure a high-quality calibration of deep photometric surveys \citep[e.g.,][and references therein]{bohlin00,holberg06,narayan16,narayan19}.

We statistically analyzed the WD locus to derive the J-PLUS absolute color calibration. The observational WD locus presents two branches, corresponding to hydrogen- (DA) and helium-dominated (DB + DC) white dwarfs \citep[e.g.,][]{holberg06,ivezic07,cfisu,gentilefusillo19,bergeron19}. These populations are evident for $\mathcal{X} = \{u, J0378, J0395, J0660\}$, where the hydrogen lines are more prominent. We performed a joint Bayesian modeling of the 11 independent $(\mathcal{X}_{\rm ISL+ZSL} - r)_0$ versus $(g-i)_0$ color-color diagrams from J-PLUS. The $r$ band was used as the absolute reference in the process. We confronted the theoretical WD locus with the observations and obtained the best parameters that modeled the observed distribution of the sources. In this process, the observational errors were accounted for.

The theoretical loci for DA and DB+DC WDs were obtained from the 3D model atmospheres presented in \citet{tremblay13} and \citet{cukanovaite18}, respectively. The high-resolution spectral models at varying gravity ($\log{\rm g} = 7,7.5,8,8.5$, and $9$ dex) were convolved with the J-PLUS filter system to obtain the theoretical WD locus. We performed a linear interpolation in the provided colors to access other gravity values during the modeling.

The assumed WD locus model has 27 parameters. The distribution in $(g-i)_0$ color was parametrized with a skew Gaussian. Its parameters were the median of the distribution ($\mu$), the dispersion ($s$), and the skew parameter ($\alpha$). The general WD population was described with two parameters: the fraction of DA white dwarfs ($f_{\rm DA}$) and the median surface gravity ($\log {\rm g}$). The offsets in each color-color diagram account for 11 parameters, parametrized with $\Delta \mathcal{C}_1$ and $\Delta \mathcal{C}_2$. These offsets impose a match between the theoretical WD locus and the observations. The offset $\Delta \mathcal{C}_2$ provides the desired $\Delta \mathcal{X}_{\rm WD}$ term in Eq.~(\ref{eq:zp}), that translates the homogenized ISL+ZSL photometry to physical units outside the atmosphere.  We defined $\Delta \mathcal{C}_1 = \Delta g_{\rm WD} - \Delta i_{\rm WD}$. This reduced the initial 20 parameters (two per color-color diagram) to 11 independent measurements. Finally, the dispersion of the WD locus in each color-color diagram amounts to 11 parameters. The diversity of white dwarf properties produces a physical dispersion in the locus after accounting for observational uncertainties. We encompassed all these physical variations in the intrinsic dispersion parameter, $\sigma_{\rm int}$. 

The modeling featured two steps. First, a simplified version of the model was run with an extra constant-density component to identify outlier WDs that are far from the theoretical WD locus. This was performed in sequence, starting from the $z$ band and moving blueward. In each color-color diagram, the outliers were identified and flagged. From the initial sample of $639$ high-quality white dwarfs in J-PLUS DR2, we identified $70$ outliers. Second, the final joint Bayesian analysis of the WD locus in the 11 color-color diagrams was performed. A total of 569 high-quality WDs were used to compute the final absolute color calibration.

We illustrate the obtained locus in Figs.~\ref{fig:wdlocus_1}, \ref{fig:wdlocus_2}, and \ref{fig:wdlocus_3}. The main results are presented and discussed in Sect.\ref{sect:wds}.

\section{Comparison with the spectroscopic standard GD 153}\label{app:gd153}
\begin{figure*}[t]
\centering
\resizebox{0.49\hsize}{!}{\includegraphics{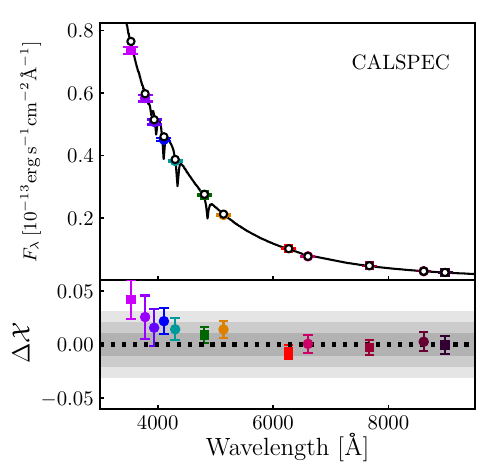}}
\resizebox{0.49\hsize}{!}{\includegraphics{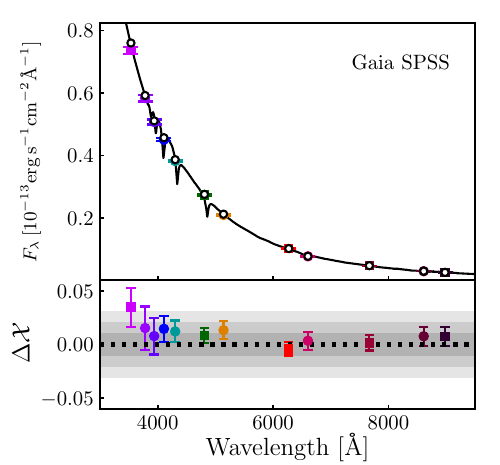}}
\caption{Comparison between the J-PLUS DR2 photometry ($\mathcal{X}_{\rm J-PLUS}$, colored points) of GD 153 and its synthetic photometry ($\mathcal{X}_{\rm standard}$, white dots) estimated from the standard spectra in CALSPEC ({\it left panel}) and {\it Gaia} SPSS ({\it right panel}). In both panels, the high-resolution standard spectra is shown with the black solid line. The magnitude difference $\Delta \mathcal{X} = \mathcal{X}_{\rm J-PLUS} - \mathcal{X}_{\rm standard}$ is shown in the lower panels. The dotted line marks a zero difference. The progressively lighter gray areas show differences of 1\%, 2\%, and 3\%, respectively.}
\label{fig:sss_gd153}
\end{figure*}
We provide an additional test to the absolute flux calibration in J-PLUS DR2. We compared the results from the final photometry with the synthetic photometry of the spectroscopic standard star GD 153. We used the reference spectra from CALSPEC\footnote{\url{https://archive.stsci.edu/hlsps/reference-atlases/cdbs/current_calspec/gd153_stiswfcnic_003.fits}} \citep{calspec,bohlin20} and from the {\it Gaia} spectro-photometric standard stars (SPSS) survey\footnote{\url{http://gaiaextra.ssdc.asi.it:8900/reduced/2/SPSSpublic/V2.SPSS003.ascii}} \citep{gaia_spss_i,gaia_spss_v}. GD 153 is one of the three calibration pillars from the Hubble Space Telescope (HST) and it was observed as part of J-PLUS DR2. The $r-$band magnitude of GD 153 in J-PLUS is $r = 13.59$ mag, so its photometry is dominated by calibration uncertainties with small photon counting errors. We note that there are other five sources in common between J-PLUS, CALSPEC, and the {\it Gaia} SPSS survey. The individual results from these sources are noisier than for GD 153, with the average result being similar to that obtained with GD 153. Moreover, the spectra of these extra sources are calibrated using the three HST pillars as references. This is the most informative and independent comparison with respect to GD 153.

The results are presented in Fig.~\ref{fig:sss_gd153}. We found a good agreement at $\lambda > 4\,500$ \AA, with differences at the 1\% level between both reference spectra and the J-PLUS photometry. The situation is less favorable at the bluer passbands, with a maximum difference of $3-4$\% in $u$. From this comparison, we can safely set the absolute flux accuracy in J-PLUS DR2 at 3\% for the $u$ band, at 2\% for $J0378$, $J0395$, $J0410$, and $J0430$; and at 1\% for $g$, $J0515$, $r$, $J0660$, $i$, $J0861$, and $z$.

The origin of the observed discrepancies in the bluer passbands can be diverse. The assumed extinction for the WD population analyzed in Appendix~\ref{app:wdlocus} could be far from the real color excess, changing the overall flux scale level. Our default procedure provided an average extinction of $E(B-V) = 0.022$ for the WD population. We tested that a zero extinction in the WD population is needed to reconcile the fluxes in the bluer passbands, but the agreement in the redder filters worsens to a 2\% level. We also tested the impact of assuming the extinction values from the 3D extinction maps \texttt{Bayestar17} and Stilism \citep{stilism}. Both maps provided $E(B-V) = 0.017$ for the WD population used for calibration, which is a small change that can account only for 0.01 mag of the discrepancy in the $u$ band.

As a final point, we stress that the comparison between the zero points in J-PLUS DR1 obtained with the WD locus and using the SDSS $u$ band as reference presented an offset of 0.04 mag \citep{clsj19jcal}. This is consistent with the offset obtained by \citet{holberg06} using WD spectra to predict photometric fluxes.
\end{appendix}

\end{document}